\documentclass[epic,eepic,fullpage,11pt]{article}
\usepackage{times}
\usepackage{amssym}

\usepackage{psfig}
\usepackage{epsfig}
\usepackage{subfigure}
\usepackage{picins}
\def\blackslug{\hcube{\hskip 1pt \vrule width 4pt height 8pt depth
 1.5pt \hskip 1pt}} \def\QED{\quad\blackslug\lower 8.5pt \null\par}     
\def\eps{\varepsilon}
\def\B{{\cal B}}
\def\T{{\cal T}}
\def\Cov{\mbox{\sl Cov}}

\def\al{\alpha}
 
\newcommand{\Fr}{Fr\'{e}chet\ }
\newcommand{\Frd}{Fr\'{e}chet distance\ }
\newcommand{\f}{\ensuremath{d_F}}
\newcommand{\wf}{\ensuremath{d_{\tilde{F}}}}
\newenvironment{proofsketch}{{\bf Proof Sketch:}}{{\rbox}}

\begin{document}
\def\polylog{{{\sl polylog }}}
\def\poly{{{\sl poly }}}
%MACROS

%
%MATH MACROS
%
\def\polylog{\mbox{\sl polylog}\,}

\newcommand{\dd}{ ,\, \ldots \, , }
\newcommand{\cd}{\,  \cdots \,  }
\newcommand{\vd}{\,  \vdots \,  }
\newcommand{\ib}{\subseteq }
\newcommand{\nmem}{\not\member}
\newcommand{\as}{\mbox{\,:=\,}}
\newcommand{\st}{\::\:}
\newcommand{\es}{\emptyset}
\newcommand{\mt}{\rightarrow}
\newcommand{\imp}{\Rightarrow}
\newcommand{\dfrac}[2]{\displaystyle{\frac{#1}{#2}}}
\newcommand{\arro}[1]{\mathop{#1}\limits^{\rightarrow}}
\newcommand{\larro}[1]{{\mathop{#1}\limits^{\longrightarrow}}}
\newcommand{\barro}[1]{{\mathop{#1}\limits^{\longleftarrow}}}
\newcommand{\dhat}[1]{{\skew4\hat\hat #1}}
\newcommand{\floor}[1]{\left\lfloor #1 \right\rfloor}
\newcommand{\bfloor}[1]{\lfloor #1 \rfloor}
\newcommand{\ceil}[1]{\left \lceil #1 \right\rceil}
\newcommand{\bceil}[1]{\lceil #1 \rceil}
\newcommand{\euclidean}{{\Bbb E}}
\newcommand{\reals}{{\Bbb R}}
\newcommand{\sphere}{{\Bbb S}}
\newcommand{\integers}{{\Bbb Z}}
\newcommand{\naturals}{{\Bbb N}}
\newcommand{\complex}{{\Bbb C}}
\newcommand{\rationals}{{\Bbb Q}}
\newcommand{\seg}[1]{\overline{#1}}
\newcommand{\li}{\item}
\newcommand{\rbox}{\begin{flushright}
        \vspace{-8mm}
        \qed
        \vspace{-1mm}
        \end{flushright}
}
\newcommand{\dstyle}{\displaystyle}
\newcommand{\sstyle}{\scriptstyle}
\newcounter{listcounter}
\newcommand{\chlist}[1]
{\renewcommand{\thelistcounter}{#1{listcounter}}}
\renewcommand{\thelistcounter}{\roman{listcounter}}
\newcommand{\descr}{\begin{list}{(\thelistcounter)}
{\usecounter{listcounter}
\setlength{\rightmargin}{0mm}}}
%
%
% New theorem environments.
%
%
\newtheorem{lemma}{Lemma}[section]
\newtheorem{theorem}[lemma]{Theorem}
\newtheorem{cor}[lemma]{Corollary}
\newtheorem{prop}[lemma]{Proposition}
\newtheorem{exer}[lemma]{Exercise}
\newtheorem{rmk}[lemma]{Remark}
\newtheorem{claim}[lemma]{Claim}
\newtheorem{fact}[lemma]{Fact}
\newtheorem{Def}[lemma]{Definition}
%
%New environments.
%
\newenvironment{pf}{{\vspace{-0.15in}\bf Proof:}}{{\rbox}}
\newenvironment{Pf}{\par\noindent{\bf Proof:}}{\mbox{}\hfill$\Box$}
\newenvironment{remark}{\noindent{\bf Remark \addtocounter{lemma}{1}
                          \hspace{-1mm}\thelemma:\hspace{2mm}}}{}
\newenvironment{remarks}{\noindent{\bf Remarks \addtocounter{lemma}{1}
                          \hspace{-1mm}\thelemma:\hspace{2mm}}}{}
\newenvironment{defn}{{\bf Definition \addtocounter{lemma}{1}
                          \hspace{-1mm}\thelemma:\hspace{2mm}}}{}
\newenvironment{Defn}[1]{{\bf Definition \addtocounter{lemma}{1}
                          \hspace{-1mm}\thelemma\ [#1]:\hspace{2mm}}}{}
\newenvironment{hint}{{\bf Hint:\hspacve{2mm}}}{}

\makeatletter
\newcounter{algo}
\def\thealgo{\@arabic\c@algo}
\def\fps@algo{tbp}
\def\ftype@algo{1}
\def\ext@algo{loa}
\def\fnum@algo{Algorithm \thealgo}
\def\algo{\@float{algo}}
\def\endalgo{\end@float}
\makeatother

\newcounter{mycounter}
%
%GENERAL MACROS
%
%\newcommand{\remarks}{{\bf Remarks: \hspace{3mm}}}
\newcommand{\Example}{{\bf Example:}}
%
%
%
%
% Macros for cross referencing.
\newcommand{\alglab}[1]{\label{alg:#1}}
\newcommand{\lemlab}[1]{\label{lemma:#1}}
\newcommand{\factlab}[1]{\label{fact:#1}}
\newcommand{\prolab}[1]{\label{prop:#1}}
\newcommand{\exelab}[1]{\label{exer:#1}}
\newcommand{\theolab}[1]{\label{theo:#1}}
\newcommand{\chaplab}[1]{\label{chap:#1}}
\newcommand{\eqlab}[1]{\label{eq:#1}}
\newcommand{\corlab}[1]{\label{cor:#1}}
\newcommand{\figlab}[1]{\label{fig:#1}}
\newcommand{\tablab}[1]{\label{tab:#1}}
\newcommand{\seclab}[1]{\label{sec:#1}}
\newcommand{\subseclab}[1]{\label{subsec:#1}}
\makeatletter
\def\remlab#1{\@bsphack\if@filesw {\let\thepage\relax
   \def\protect{\noexpand\noexpand\noexpand}%
\xdef\@gtempa{\write\@auxout{\string
           \newlabel{rem:#1}{{\thelemma}{\thepage}}}}}\@gtempa
            \if@nobreak \ifvmode\nobreak\fi\fi\fi\@esphack}
\def\deflab#1{\write\@auxout{\string
        \newlabel{def:#1}{{\thelemma}{\thepage}}}}
\makeatother

\newcommand{\lemref}[1]{Lemma~\ref{lemma:#1}}
\newcommand{\algref}[1]{Algorithm~\ref{alg:#1}}
\newcommand{\factref}[1]{Fact~\ref{fact:#1}}
\newcommand{\proref}[1]{Proposition~\ref{prop:#1}}
\newcommand{\exeref}[1]{Exercise~\ref{exer:#1}}
\newcommand{\theoref}[1]{Theorem~\ref{theo:#1}}
\newcommand{\corref}[1]{Corollary~\ref{cor:#1}}
\newcommand{\figref}[1]{Figure~\ref{fig:#1}}
\newcommand{\tabref}[1]{Table~\ref{tab:#1}}
\newcommand{\Figref}[1]{Figure~\ref{fig:#1}}
\newcommand{\chapref}[1]{Chapter~\ref{chap:#1}}
\newcommand{\secref}[1]{Section~\ref{sec:#1}}
\newcommand{\subsecref}[1]{Section~\ref{subsec:#1}}
\newcommand{\remref}[1]{Remark~\ref{rem:#1}}
\newcommand{\defref}[1]{Definition~\ref{def:#1}}
\newcommand{\mchapref}[1]{\ref{chap:#1}}
\newcommand{\msecref}[1]{\ref{sec:#1}}
\newcommand{\msubsecref}[1]{\ref{subsec:#1}}
\newcommand{\mtheoref}[1]{\ref{theo:#1}}
\newcommand{\mfigref}[1]{\ref{fig:#1}}
\newcommand{\eqref}[1]{(\ref{eq:#1})}
%
%
%
% Macros for generating table of contents
\newcommand{\secadd}[2]{\addcontentsline{toc}{section}{
                \protect\numberline{\msecref{#1}}{#2}}}
\newcommand{\subsecadd}[2]{\addcontentsline{toc}{subsection}{
                \protect\numberline{\msubsecref{#1}}{#2}}}
\newcommand{\chapadd}[2]{\addcontentsline{toc}{chapter}{
                \protect\numberline{\mchapref{#1}}{#2}}}
\newcommand{\figadd}[2]{\addcontentsline{lof}{figure}{
                \protect\numberline{\mfigref{#1}}{#2}}}

%%%%%%%%%%%%%%%%%%%%%%%%%%%%%%%%%%%%%%%%%%%%%%%%%%%%%%%%%
%Producing table of contents entries without
%numbers.
%%%%%%%%%%%%%%%%%%%%%%%%%%%%%%%%%%%%%%%%%%%%%%%%%%%%%%%%%
\newcommand{\nsecadd}[1]{\addcontentsline{toc}{section}{#1}}
\newcommand{\nsubsecadd}[1]{\addcontentsline{toc}{subsection}{#1}}
\newcommand{\nchapadd}[1]{\addcontentsline{toc}{chapter}{#1}}
%
%
% Macros for page style.
{\catcode`\@=11
\gdef\setft#1#2#3{%
\def\@oddfoot{
{\setbox0=\hbox{#1}
\setbox1=\hbox{#3}
\ifdim\wd0>\wd1
\dimen0=\wd0
\box0\hfil#2\hfil\hbox to\dimen0{\hfil\hfil\box1}
\else \dimen0=\wd1
\hbox to\dimen0{\box0\hfil}\hfil#2\hfil\box1\fi
}}}}

{\catcode`\@=11
\gdef\sethd#1#2#3{%
\def\@oddhead{\vbox{\hbox to\hsize{{#1}\hfil{#2}\hfil{#3}}%
\vspace{0.06in}%
\hbox to \hsize{\hrulefill}\vspace*{-0.09in}}}
\def\@evenhead{\@oddhead}
        }
}

%Macro for sections.
\def\mysec#1{\setcounter{equation}{0}
\section{#1}\markright{#1}}

\def\mysecn#1{\setcounter{equation}{0}
\section*{#1}\markright{#1}}

\def\myplsec#1{\setcounter{equation}{0}
                \section*{#1}\mark{#1}}

\def\thebibliography#1{\mysecn{References}\list
{[\arabic{enumi}]}{\settowidth\labelwidth{[#1]}\leftmargin\labelwidth
 \advance\leftmargin\labelsep
 \usecounter{enumi}}
 \def\newblock{\hskip .11em plus .33em minus .07em}
 \sloppy\clubpenalty4000\widowpenalty4000
 \sfcode`\.=1000\relax}
\let\endthebibliography=\endlist
  %%%%%%%%%%%%%%%%%%%%%%%%%%%%%%%%%%%%%%%%%%%%%%%%%%%%%%%%%%%%%%%%%%%%%%%%
  %%  Some code to insert marginal comments and number them in a paper. %%
  %%  By default marginal comments are ignored.  If the command         %%
  %%  \withcomplaints is called, complaints are numbered, the number    %%
  %%  between dashes appears on the tex raised in a 0 width box so it   %%
  %%  does not affect the layout.                                       %%
  %%%%%%%%%%%%%%%%%%%%%%%%%%%%%%%%%%%%%%%%%%%%%%%%%%%%%%%%%%%%%%%%%%%%%%%%

\def\complaint#1{}
\def\withcomplaints{
%\addtolength{\oddsidemargin}{-1.4cm}
%\addtolength{\evensidemargin}{-1.4cm}
\newcounter{mycomplaints}
\def\complaint##1{\refstepcounter{mycomplaints}%
\ifhmode%
\unskip%
{\dimen1=\baselineskip \divide\dimen1 by 2 %
\raise\dimen1\llap{\tiny -\themycomplaints-}}\fi%
\marginpar{\tiny [\themycomplaints]: ##1}}%
}

\newcounter{printertype}
\setcounter{printertype}{0}
\def\figprint#1{
        \ifcase \theprintertype

                \begin{center}
                 \input{#1}
                \end{center}
              \or
                 \centerline{\psfig{figure=#1.ps}}
              \else
                 \vspace*{1in}
        \fi}

\makeatletter
\long\def\@myfootnotetext#1{\insert\footins{\footnotesize
    \interlinepenalty\interfootnotelinepenalty 
    \splittopskip\footnotesep
    \splitmaxdepth \dp\strutbox \floatingpenalty \@MM
    \hsize\columnwidth \@parboxrestore
   \edef\@currentlabel{\csname p@footnote\endcsname\@thefnmark}\@makemyfntext
    {\rule{\z@}{\footnotesep}\ignorespaces
      #1\strut}}}

\def\myfootnotetext{\@ifnextchar
[{\@xfootnotenext}{\xdef\@thefnmark{\thempfn}\@myfootnotetext}}

\long\def\@makemyfntext#1{\parindent 5mm #1}
\makeatother

%%%%%%%%%%%%%%%%%%%%%%% bibmacros %%%%%%%%%%%%%%%%%%%%%%%
%%%%%%%%%%%%%%%%%%%%%%%%%% Conferences %%%%%%%%%%%%%%%%%%
\def\th{$^{\mbox{\footnotesize th}}$}
\newcommand{\scg}[1]{{\it Proceedings #1\th Annual Symposium
on Computational Geometry}}
\newcommand{\focs}[1]{{\it Proceedings #1\th Annual
IEEE Symposium on Foundations of Computer Science}}
\newcommand{\stoc}[1]{{\it Proceedings #1\th Annual
ACM Symposium on Theory of Computing}}
\newcommand{\icalp}[1]{{\it Proceedings #1\th
International Colloquium on Automata, Languages and Programming}}
\newcommand{\soda}[1]{{\it Proceedings #1\th Annual
ACM-SIAM Symposium on Discrete Algorithms}}
\newcommand{\swat}[1]{{\it Proceedings #1\th Scand. 
Workshop on Algorithms Theory}}
\newcommand{\wads}[1]{{\it Proceedings #1\th Workshop
on Algorithms and Data Structures}}
\newcommand{\cccg}[1]{{\it Proceedings #1\th Candian
Conf.\ Computational Geometry}}

%%%%%%%%%%%%%%%%%%%%%%%%% Journals %%%%%%%%%%%%%%%%%%%%%%%%%

\newcommand{\aaa}{{\it Acta Arithmetica}}
\newcommand{\aia}{{\it Acta Informatica}}
\newcommand{\algc}{{\it Algorithmica}}
\newcommand{\ajm}{{\it American J.\ Mathematics}}
\newcommand{\aujm}{{\it Australian J.\ Mathematics}}
\newcommand{\ama}{{\it Acta Mathematica}}
\newcommand{\amh}{{\it Acta Mathematica Acad.\ Sci.\ Hung.}}
\newcommand{\amm}{{\it American Mathematical Monthly}}
\newcommand{\asm}{{\it Acta Scientiarum Mathematicarum (Szeged)}}
\newcommand{\bams}{{\it Bull. American Mathematical Society}}
\newcommand{\blms}{{\it Bull. London Mathematical Society}}
\newcommand{\cacm}{{\it Communications of  ACM}}
\newcommand{\cg}{{\it Computer Graphics\/}}
\newcommand{\cpam}{{\it Communications of  Pure Applied Mathematics}}
\newcommand{\cgta}{{\it Computational Geometry:
                        Theory and Applications}}
\newcommand{\cgip}{{\it Computer Vision, Graphics
                        and Image Processing}}
\newcommand{\cjb}{Colloq.\ Mathematical Society J\'{a}nos Bolyai}
\newcommand{\cjm}{{\it Canadian J.\ Mathematics}}
\newcommand{\cmb}{{\it Canadian Math.\ Bulletin}}
\newcommand{\comb}{{\it Combinatorica}}
\newcommand{\colm}{{\it Colloquium Mathematicum}}
\newcommand{\dcg}{{\it Discrete and Computational Geometry}}
\newcommand{\dam}{{\it Discrete Applied Mathematics}}
\newcommand{\dcm}{{\it Discrete Mathematics}}
\newcommand{\djm}{{\it Duke J.\ Mathematics}}
\newcommand{\ejc}{{\it European J.\ Combinatorics}}
\newcommand{\gc}{{\it Graphs and Combinatorics}}
\newcommand{\gtc}{{\it Graph Theory and Combinatorics}}
\newcommand{\ieeec}{{\it IEEE Transactions on Computers}}
\newcommand{\ijcga}{{\it International J.\ of Computational 
                         Geometry and Applications}}
\newcommand{\inm}{{\it Inventiones Mathematicae}}
\newcommand{\ijm}{{\it Israel J.\ Mathematics}}
\newcommand{\ipl}{{\it Information Processing Letters}}
\newcommand{\jacm}{{\it J. Assoc.\ Comput.\ Mach.}}
\newcommand{\jalg}{{\it J. Algorithms}}
\newcommand{\jams}{{\it J. American Mathematical Society}}
\newcommand{\jlms}{{\it J. London Mathematical Society}}
\newcommand{\jcss}{{\it J. Computer and Systems Sciences}}
\newcommand{\jct}{{\it Journal of Combinatorial Theory}}
\newcommand{\jsc}{{\it J. Symbolic Computation}}
\newcommand{\ma}{{\it Mathematische Annalen}}
\newcommand{\matha}{{\it Mathematika}}
\newcommand{\mcl}{{\it Machine Learning}}
\newcommand{\mtam}{{\it Magyar Tud.\ Akad.\ Mat.\ Kutat\'o
                                Int\'ezet K\"ozlem\'enyei\/}}
\newcommand{\pami}{{\it IEEE Trans.\ Pattern Anal.\ and
                                Mach.\ Intell.}}
\newcommand{\pams}{{\it Proc. American Mathematical Society}}
\newcommand{\pjm}{{\it Pacific J.\ Mathematics}}
\newcommand{\plms}{{\it Proc. London Mathematical Society}}
\newcommand{\pmh}{{\it Periodica Mathematica Hungarica}}
\newcommand{\pr}{{\it Pattern Recognition}}
\newcommand{\prl}{{\it Pattern Recognition Letters}}
\newcommand{\soc}{{\it SIAM J. Computing}}
\newcommand{\ssmh}{{\it Studia Scientiarum Mathematicarum Hungarica}}
\newcommand{\tcs}{{\it Theoretical Computer Science}}
\newcommand{\tams}{{\it Trans.\ American Mathematical Society}}

%%%%%%%%%%%%%%%%%%%%% Misc. %%%%%%%%%%%%%%%%%%%%%
\newcommand{\mas}{manuscript}
\newcommand{\trp}{Technical Report}
\newcommand{\dcs}{Dept. Computer Science}
\newcommand{\dpm}{Dept. of Mathematics}
\newcommand{\nyu}{New York University}
\newcommand{\ipn}{in preparation}
\newcommand{\pcn}{personal communication}

%%%%%%%%%%%%%%%%%%%%%% Author abbrevs %%%%%%%%%%%%%%%%%%%%
\newcommand{\chaz}{B. Chazelle}
\newcommand{\chazelle}{B. Chazelle}
\newcommand{\edels}{H. Edelsbrunner}
\newcommand{\guib}{L. Guibas}
\newcommand{\guibas}{L. Guibas}
\newcommand{\overm}{M. Overmars}
\newcommand{\seid}{R. Seidel}
\newcommand{\shar}{M. Sharir}
\newcommand{\sharir}{M. Sharir}
\newcommand{\welzl}{E. Welzl}
\newcommand{\clark}{K. Clarkson}
\newcommand{\mato}{J. Matou\v{s}ek}
\newcommand{\Erdos}{P. Erd\H{o}s}
\newcommand{\szem}{Szemer\'{e}di}

%%% Local Variables: 
%%% mode: latex
%%% TeX-master: t
%%% End: 

\newtheorem{problem}{Problem}[section] 
\newcommand{\probref}[1]{Problem~\ref{prob:#1}} 
\newcommand{\problab}[1]{\label{prob:#1}} 
\newcommand{\claimref}[1]{Claim~\ref{claim:#1}} 
\newcommand{\claimlab}[1]{\label{claim:#1}}

\sethd{{\sc \firstmark}}{}{{\rm \thepage}} %%
%\setft{{\it \hspace{-4mm} On Segment Matching\hspace{-30mm}}}{}{{\hspace{-30mm}\it \today}}
%\setft{{\it \hspace{-4mm} Segment Matching\hspace{-30mm}}}{}{{\hspace{-30mm}\it \today}}

\title{
Pattern Matching for Sets of Segments\thanks{A full version of this paper
  can be found at http://graphics.stanford.edu/$\sim$alon/papers/seg\_match.ps.gz}
}

\author{
Alon Efrat\thanks{ 
Computer Science Department, Stanford University,
\textbf{Email: }{\sl alon@cs.stanford.edu}.
Work supported in part by a Rothschild Fellowship and by
DARPA contract DAAE07-98-C-L027}
\and 
Piotr Indyk\thanks{ 
Computer Science Department, Stanford University.
\textbf{Email: }{\sl indyk@cs.stanford.edu}.} 
\and
 Suresh Venkatasubramanian\thanks{AT\&T Labs -- Research. \textbf{Email:
     }\textsl{suresh@research.att.com}.} 
}

\def\frs{Fr\'echet} 
\def\TP{{\sl TP}\ } 
\def\S{{\cal S}} 
\def\maxs{\max(\sum_\S(\cdot))}
\def\sum{\mbox{\sl sum}}
\def\F{{\cal F}} 
\def\epsilon{\varepsilon} 
\def\Pos{{\sl Pos}}
\def\Neg{{\sl Neg}}
\long\gdef\boxit#1{\vspace{5mm}\begingroup\vbox{\hrule\hbox{\vrule\kern3pt
\vbox{\kern4pt#1\kern3pt}\kern3pt\vrule}\hrule}\endgroup}
\newcommand{\qed}{\mbox{}\hspace*{\fill}\nolinebreak
 \mbox{$\rule{0.7em}{0.7em}$}}

\newcommand{\comment}[1]{}

\begin{titlepage}
\def\thepage{}
\maketitle  

\begin{abstract} 
   
In this paper we present algorithms for a number of 
 problems in geometric pattern matching
where the input consist of a collections of segments in the plane.
Our work consists of two main parts. In the first, we address problems and
measures that relate to collections of orthogonal line segments in the
plane. Such collections arise naturally from problems in mapping
buildings and robot exploration. 

We propose a new measure of segment
similarity called a 
\emph{coverage measure}, and present efficient algorithms for 
maximising this  measure between sets of axis-parallel segments under
translations. Our algorithms run in time $O(n^3\polylog n)$ in the general
case, and run in time $O(n^2\polylog n)$ for the case when all segments
are horizontal. In addition, we show that when restricted to translations
that are only vertical, the Hausdorff distance between two sets of
horizontal segments can be computed in time roughly $O(n^{3/2}\mbox{\sl
  polylog\,}n)$.  These algorithms form  significant improvements
 over the general algorithm of Chew et al.\  that
takes time  $O(n^4  \log^2 n )$. 

In the second part of this paper we address the problem of matching
polygonal chains. We study the well known \Frd, and present the first
algorithm for computing the \Frd\  under general translations. Our methods
also yield algorithms for computing a generalization of the \Fr distance,
and we also present a simple approximation algorithm for the \Frd\ that
runs in time $O(n^2\polylog n)$.   
   
\end{abstract} 
%\begin{center}
%\boxit{\Large{10:36 PM Jul 6}}
%\end{center}
\end{titlepage}

\mysec{Introduction} 

Traditionally, geometric pattern matching employs as a measure of
similarity the Hausdorff distance h(A,B), defined as $h(A,B) = 
\max_{p \in A}~\min_{q \in B}~ d(p,q)$ for two point sets $A$ and
$B$. However, when the patterns to be matched are line segments or curves
(instead of points), this measure is less than satisfactory. It has been
observed that measures 
like the Hausdorff measure that are defined on point sets  are ill-suited as
measures of curve similarity, because they destroy the continuity inherent
in continuous curves. 

This paper addresses problems in geometric pattern matching where the
inputs are sets of line segments. Our work consists of two main parts; in
the first part we consider the problem of matching (under translation)
segments that are axis-parallel (i.e either horizontal or vertical), and
in the second we consider the problem of matching polygonal chains under
translation. We study two different measures in this context; the first is
a novel measure called the \emph{coverage measure}, which captures the
similarity between orthogonal segments that may partially overlap with
one another. The other is the
well known \Fr distance, first proposed by Maurice \Fr in 1906 as a
measure of distance between distributions,  which  has often been
referred to as a natural measure of curve
similarity~\cite{AG92,BW98,WN94}.  We discuss each measure in 
detail below.

\subsection{Mapping and orthogonality}
\label{ssec:map}
The motivation for considering instances of pattern matching where the
input line segments are orthogonal comes from the domain of
\emph{mapping}, in which a robot is required to map the underlying
structure of a building by moving inside the building, and ``sensing'' or
``studying'' its environment.
 
In one such mapping project at the Stanford Robotics laboratory\footnote{
The interested reader can find more information at the URL {\sl
underdog.stanford.edu} } the robot is equipped with a laser range finder
which supplies the distance from the robot to its nearest neighbor in a
dense set of directions in a horizontal plane. We call the resulting
distances map a {\em picture}.
%once the robot 
%is static, 
\figref{robot}(a)  shows  the  robot used at Stanford 
for this purpose    the laser range finder installed on the robot. 

During the mapping process, the robot must merge into a single  map the series
of pictures that it captures from different locations in the building.
\vspace{-0.2in}
\newlength{\figwidth} \setlength{\figwidth}{2in} 
\begin{center}
\begin{figure}[h]
\subfigure[]{\epsfig{file=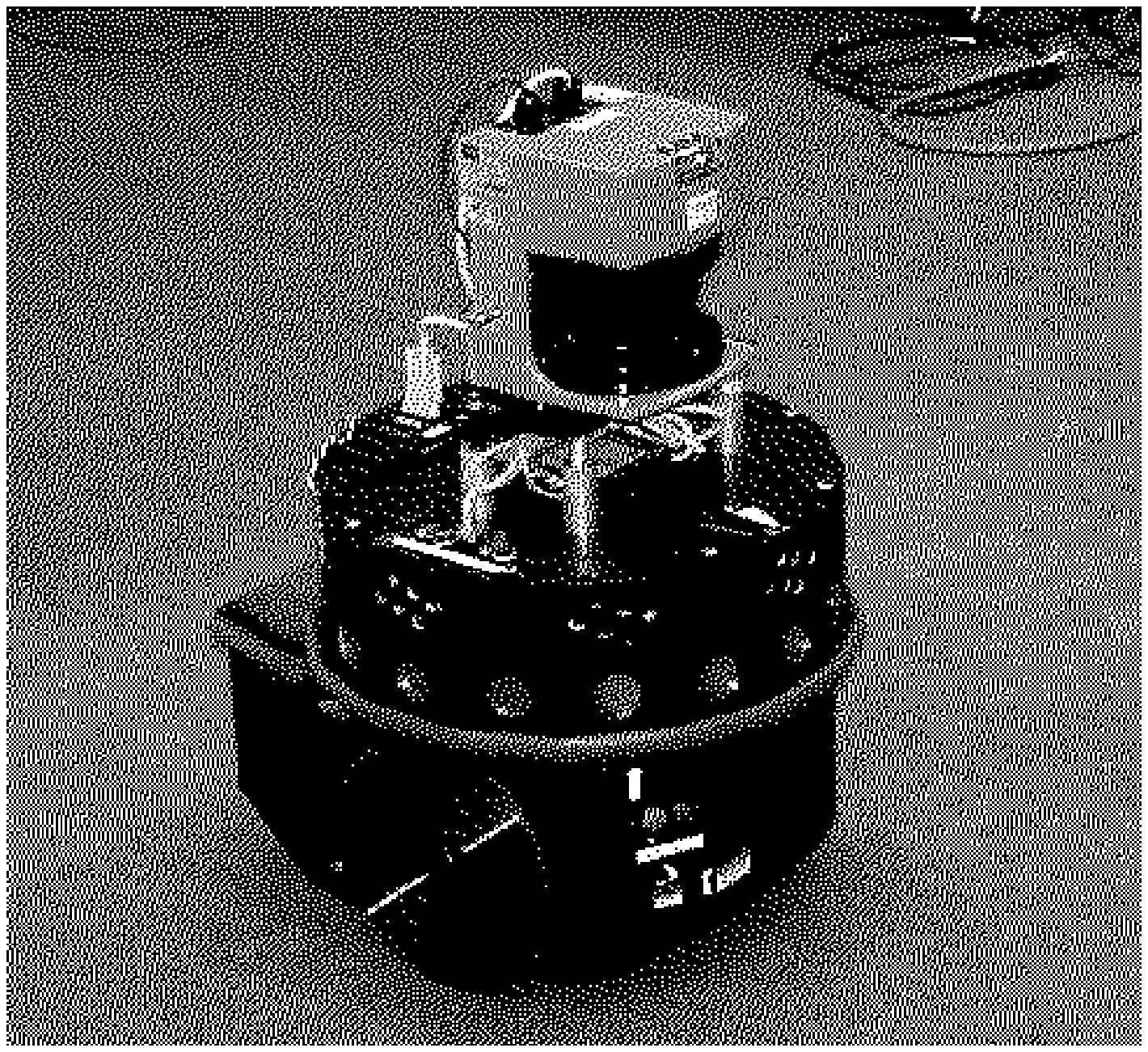,width=\figwidth}}
\subfigure[]{\epsfig{file=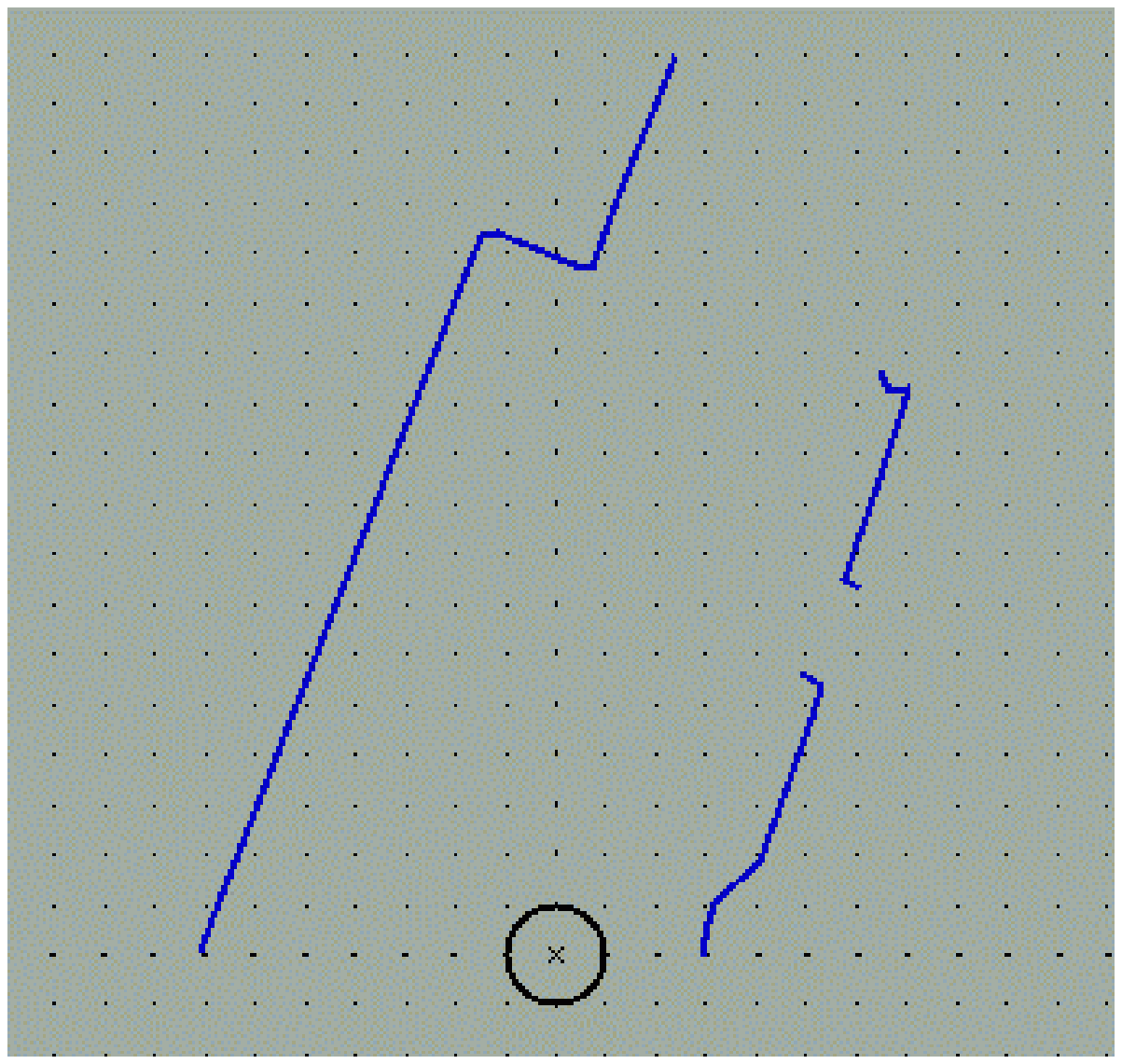,width=\figwidth}} 
\subfigure[]{\epsfig{file=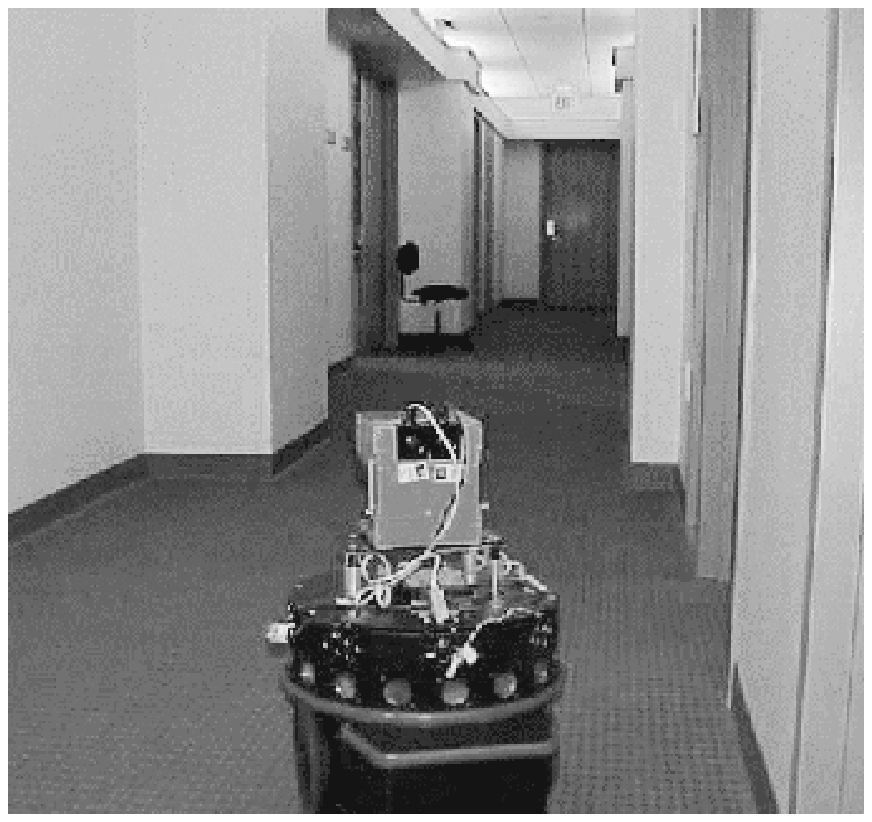,width=\figwidth}
} \caption{\figlab{robot} Left: The robot, and the  laser range finder
installed on it. Middle: Typical ``picture'' obtained by the robot of a
corridor (after segmentation).  Right: The corridor itself} 
\end{figure}
\end{center}  
\vspace{-0.2in}
Since the dead reckoning of the robot is not very accurate, it cannot rely
solely on its motion to decide how the pictures are placed together. Thus,
we need a matching process that can  align (by using overlapping regions)
the different pictures taken from different points of the same
environment. In addition, we need to determine 
whether the robot has returned to a point already visited. We make the
reasonable assumption that buildings walls are almost always either
orthogonal or parallel to each other, and that these walls are frequently
by far the most dominant objects in the pictured. This is especially
significant in the case that the robot is inside a corridor, where there
is a lack of detail needed for good registration. In some cases
most of the picture consists merely of two walls with a small number of
other segments. See \figref{robot}(b),(c) for a typical picture and the real
region that the laser range finder  senses. 

This application suggests the study of matching sets of horizontal and
vertical segments. Observe that we may restrict ourself to alignments 
under {\em translation}, as it is easy to find the correct rotation for
matching sets of  orthogonal segments. 
Formally, let $A=\{a_1\dots a_n\}$ and $B=\{b_1,\dots b_n\}$ be two sets
of orthogonal line segments in the plane, and let $\eps$ be a given parameter.  A
point $p$ of a  horizontal (resp.\ vertical ) segment  $a\in A$ is 
{\em covered} if there is a point of a horizontal (resp.\ vertical) 
segment $b\in B$ whose distance from $p$ is $\leq \eps$, where the
distance is measured using the $\ell_\infty$ norm.  Let $w(A,B)$ denote the
collection of sub-segments of $A$ consisting of covered points.  Let
$\Cov(A,B)$ be the total length of the segments of $w(A,B)$.  The {\em
maximum coverage problem} is to find a translation $t^*$ in the
translation plane ({\em \TP}) that maximizes $\Cov(t) =
\Cov(t+A, B)$. To the best of our knowledge, this measure is novel. 

The coverage measure is especially relevant in the case 
of long segments e.g.\ inside a corridor, when we might be interested in 
partially matching portions of long segments to portions of other
segments.
\vspace{-0.2in}
\paragraph{Our Results}
In \secref{coverage} we present an  algorithm that solves the Coverage
problem between   
sets of axis-parallel segments in time $O(n^{3}\log^2 n)$  and the
Coverage problem between horizontal segments in time $O(n^2\log n)$
Note that the known algorithms  for matching
arbitrary sets of line segments are much slower. For example, the best known
algorithm for finding a translation that minimizes the Hausdorff Distance
between two sets of $n$ segments in the plane runs in time $O(n^4 \log ^2
n  )$ \cite{AST,cghkkk-gpmem-93}. We also show that the that the
combinatorial complexity of the Hausdorff matching   
between segments is $\Omega(n^4)$, even if all segments are {\em
  horizontal}. This strengthens  the  bounds shown by Rucklidge
\cite{Rucklidge-lower-bound}, and demonstrates that our algorithms,  much
like the algorithms of \cite{ChewKedem,CDEK}  are able to avoid having to
examine each cell of $\F$ individually.  Note that all our results extend
to the case when segments are \emph{weighted} and the coverage is now a
weighted sum of interval lengths.

In Section~\secref{vertical-translation}
we  consider the related problem of matching horizontal segments under
vertical translations (under the Hausdorff measure). It has been observed
that if horizontal translations are allowed, then this problem is
3SUM-hard~\cite{Sariel}, indicating that finding a sub-quadratic algorithm
may be hard. However, we present an algorithm running in time $O(n^{3/2}
\max\{ \log^c M, \log^c n, 1/\epsilon^c)) \} $, for some fixed constant
$c$, which is sub-quadratic in most cases. Here, $M$ denotes the ratio of
the diameter to the closest pair of points in the sets of segments (where
pairs of points must lie on different segments).

%A question that addresses the intrinsic complexity of the coverage problem
%is:  what is the complexity of the region $\F$ in the translation plane  
%of all translation $t$ for which $\Cov(t+A, B)$ reaches some value $L$. 
%This question is   interesting because  for many geometric 
%pattern matching problems the best known algorithms explicitly computes all
%acceptable solutions and there are no known methods for computing just one
%solution. The  question of the complexity of $\F$  is clearly
%as hard  as the equivalent question for the Hausdorff distance, since we
%can use the coverage to answer  
%the question for the Hausdorff distance. Rucklidge \cite{Rucklidge-lower-bound}
% presented a construction of segments for which the complexity of 
%$\F$ is $\Theta(n^4)$ (for the Hausdorff measure). However, his
%construction uses segments of arbitrary  
%directions. In Section~\secref{lower-bound} we show 

\subsection{The \Frd}
\label{ssec:frechet}

In the second part of the paper, we consider measures for matching
polygonal chains under the \Fr distance. 
Let us define a curve as a continuous mapping $P : [a,a'] 
\rightarrow \reals^2$. The \Frd between two curves $P$ and $Q$, $d_F(P  
,Q)$ is defined as: 
\[ d_F(P,Q) = \inf_{\alpha,\beta} \max_{t \in [0,1]}
\|f(\alpha(t))-g(\alpha(t))\| \]  
where $\alpha,\beta$ range over continuous increasing functions from
$[0,1]\rightarrow [a,a']$ and $[0,1]\rightarrow [b,b']$ respectively.

Alt and Godau proposed the first algorithm for computing the \Frd\ between
two polygonal chains (with no transformations). Their method is elegant
and simple, and runs in time $O(pq)$, where $p$ and $q$ are the number of
segments in the two polygonal chains. 
In his Ph.D thesis~\cite{G98}. Michael Godau presents an extensive study
of the complexity of computing the \Fr distance. He shows that
computing the \Frd\ between two simplicial objects is NP-hard, for any
dimension $d \ge 3$.  %He also gives a more general proof of the fact that
%for convex objects, the \Frd\ and the Hausdorff distance coincide; the
%original proof was given in~\cite{ABGW90}.

Although the \Fr distance is a natural measure for curve similarity,
its applicability has been limited by the fact that no algorithms exist to
minimise the \Fr distance between curves under various transformation
groups. Prior to our work, the only result on computing the \Frd\ under
transformations was presented by Venkatasubramanian~\cite{V99}. He computes 
$\min_{t \in TP_x}  d_F(P, Q+t) \le \epsilon$, where $TP_x$ is the set of
translations along a fixed direction, in time $O(n^5\mbox{\sl
  polylog\,}n)$ (where $n=p+q$). In fact, our methods can be viewed as a
generalization of his methods and can be used to solve his problem in the same 
time bound.
\vspace{-0.15in}
\paragraph{Our Results}
In \secref{frechet} we present the first algorithm for computing the \Frd\ between two
polygonal chains minimized under translations\footnote{Actually, we solve
  the decision version of the problem: For a given $\epsilon$, determine whether
  $\min_{t \in TP}  d_F(P, Q+t) \le \epsilon$.}. The algorithm is based on
a reduction to a dynamic graph reachability problem; its running time is
$O(n^{10}\polylog n)$. 

If we drop the restriction that the functions 
 $\alpha, \beta$ must be  increasing, we obtain 
 a measure that we call the \emph{weak} \Frd, denoted by 
$\wf$. Our methods can be used to decide whether 
$\min_{t \in TP}  \wf(P,Q+t) \le \epsilon$; in this case, the underlying
graph is undirected, yielding an algorithm that runs in time
$O(n^4\polylog n)$.  

With the exact algorithms being rather expensive, it is natural to ask whether
approximations can be obtained efficiently. A simple observation shows
that we can obtain an $(\epsilon,\beta)$-approximation to the \Frd\ under
translations in time $O(n^2\poly(\log n,1/\beta))$.

%--------------------------------- 
%\pagestyle{empty} 
\mysec{Maximum Coverage Among Sets Of Segments}\seclab{coverage} 

Let $A=\{a_1\dots a_n\}$ and $B=\{b_1,\dots b_n\}$ be two sets of
axis-parallel line segments in the plane, and let $\eps$ be a given
parameter.  Recall the coverage measure $\Cov(A,B)$ as defined in the
introduction. 
%-----------------------
%the problem of  {\em coverage} problem, as defined in the introduction. 
% A point $p$
%of a segment $a\in A$ is {\em covered} if there is a point of a segment
%$b\in B$ whose distance from $p$ is $\leq \eps$, where the distance is
%measured using the $\ell_infty$ norm.  Let $w(A,B)$ denote the collection of
%sub-segments of $A$ consisting of covered points.  Let $\Cov(A,B)$ be the
%total length of the segments of $w(A,B)$.  The {\em maximum coverage
%problem} is to find a translation $t^*$ in the translation plane
%(abbreviated {\em \TP}) that maximises $\Cov(t) = \Cov(t+A, B)$ under all
%translations $t$.
%-----------------
% This problem is clearly at least as hard as the Hausdorff
%distance problem, since the maximum coverage  might be obtained where all segments of
%$A$ are covered.  Yet we are able to give an $O(n^2 \log n)$-time
%algorithm for the problem, for the case when all the segments are
%horizontal, and an $O(n^3\polylog n)$ time algorithm 
%in the case that the sets consists of both horizontal and vertical segments.
%
%-------------------------------

\subsection{Computing coverage with axis-parallel segments} 
\subseclab{horizontal+vertical}
We first consider the case that the sets $A$ and $B$ consists of 
both horizontal and vertical segments. 
Let $A^h$ (resp.\ $B^h$) be a set of $n$ horizontal segments and let $A^v$
(resp.\ $B^v$) be a set of $n$ vertical segments. Let  $\eps$ be a
given parameter.
Let $A=A^h\cup A^v$ and let $B=B^h\cup B^v$.  Let $\Cov(t+A, B) =
\Cov(t+A^h, B^h)+ \Cov(t+A^v, B^v)$.

We first need the following  lemma, whose 
proof is deferred to Appendix~\ref{app:pf1}.
Let $\S=\{s_1\dots s_m\}$ be a set of non-vertical segments in
$\reals^2$. For each segment $s_i\in\S$ we define the functions
$s_i(x)\rightarrow \reals$\ as follows: For every $x\in\reals$, $s_i(x)$
is the $y$-coordinate of the intersection point of $s$ and the vertical
line passing through $x$, if such an intersection point exists. We set
$s_i(x)$ to be $0$ otherwise. Let $\sum_\S(x) = \Sigma_{i=1}^m s_i(x)$, and let $\maxs =
\max_{x\in\reals}\sum_\S(x)$.  Furthermore, let $T=T(\tau)$ be a subset of
$\S$ consisting of horizontal segments that can move vertically at
constant speed i.e  the $y$-coordinates of the endpoints of each $s_i\in
T$ are given by $y = a_i\tau+b_i$. 

\begin{lemma}\lemlab{zzz2}
Given a set of non-vertical segments $S$ with a subset $T$ of horizontal
moving segments, we can maintain $\maxs$ under segment insertions or
deletions in  amortized time $O(\sqrt |\S|)$ per operation. In addition,
we can maintain $\maxs$ under a \emph{time-decreasing} step ($\tau
\leftarrow \tau - \Delta$) in $O(1)$ time. 
\end{lemma}

\begin{theorem}\theolab{combined} 
We can find a translation $t$ that maximizes $\Cov(t+A, B)$
 in time $O(n^3 \log^2 n)$, where $n= |A|+ |B|$ 
\end{theorem}  
%------------------
\begin{pf}
The proposed algorithm is a line-sweep algorithm, with the sweep line 
moving from top to bottom.
For a segment $b_i\in B$ let $b_i^+$ denote the rectangle consisting of
all points whose $\ell_infty$ distance from $b_i$ is at most $\eps$.  Let
$B^+$ denote the union $\bigcup_{i=1}^n b_i^+$.  Note that any two
rectangles $b_i^+, b_j^+$ intersect in at most two points, so by
\cite{KLPS} the complexity of the boundary of $B^+$ is $O(n)$. Consider
$E=\{p_1\dots p_{2n} \}$, the set of the $2n$ endpoints of the segments of
$A$. Define the {\em layer} $L_i =B^+-p_i$, which is the region in the
\TP\ of all translations 
$t$ that shift $p_i$ into $B^+$ i.e $t+p_i\in B^+$. 
  Let $\B^h$ (resp.\ $\B^v$) be the collection of layers
created by the horizontal (resp.\ vertical) segments of $A$.
As the line sweep traverses the translation plane from top to bottom, we
encounter events where $\ell$ intersects a horizontal boundary segment of
either $\B^h$ or $\B^v$.

\noindent{\bf Horizontal Boundaries Of $\B^h$: }
% We assign to each point $t$ of the line sweep $\ell$ the value $x_t$,
% where $x_t$  is the orthogonal projection of $t$ on the $x$-axis, 
%and
Let $\Cov(x) : \reals \rightarrow \reals$ be the value of $\Cov(t+A^h,B^h)$, where 
$t$ is the point on $\ell$ vertically above $x$. Consider the contribution
to $\Cov(t+A^h, B^h)$  from the interaction between the  segments  $a\in
A^h, b\in B^h$.  
This contribution  to the function 
consists of a piecewise linear function, consists of five segments: 
It is zero for value of $x$ which are very far from the regions of interaction 
between $a$ and $b$, it is a constant that equals the minimum of the length of 
$a$ and $b$ when $x$ is near the region of intersection, and it consists of two 
 segments of slopes are $1$ and $-1$, 
connecting these  segments. These segments exist for all instances of the
line sweep where  its  horizontal distance to the boundary of the
rectangle of $B_j$ corrsponds to $a_i$ is $\leq \eps$.  
There are $O(n^2)$ update operations, and each update can be processed in
$O(n\log^2n)$ time from Lemma~\lemref{zzz2}. 

\noindent{\bf Horizontal Boundaries Of $\B^v$. }
For two vertical segments $a_i\in A, b_j\in B$, let $\T_{ij}$ be the set
of translations for which the horizontal distance from $a_i$ to $b_j$ is
at most $\eps$.  Assume w.l.o.g that $|a_i|>|b_j|$. Let $UP_{ij}$ denote
all translations $t$ for which the  upper endpoint of $a_i$ is covered by
$t+b_j$, (i.e. its distance from some point of $t+b_j$ is at most $\eps$)
but the lower endpoint of $a_i$ is not covered. Similarly, let $DOWN_{ij}$
denote  all translations $t$ for which the lower endpoint of $a_i$ is
covered by $t+b_j$,  but the upper  endpoint of $a_i$ is not covered and
let $MID_{ij}$ denote all translations $t$ for which  both endpoints of
$a_i$ are covered. 
 
Thus $\Cov(t+a_i, b_j)$ is zero when 
$t\notin UP_{ij}\cup MID_{ij}\cup DOWN_{ij}$, 
  $\Cov(t+a_i, b_j)$  is a constant when $t\in MID_{ij}$,
 and it is  a decreasing (resp.\ increasing) linear function that depends only on the
 $y$-coordinate of $t$ when $t\in UP_{ij}$  (resp.\    $t\in DOWN_{ij}$). 
Therefore,  we can represent the contribution of $a_i$ and $b_j$ to $\Cov(a_i,
t+b_j)$ by a horizontal segment $u_{ij}(\tau)$ of length $2\epsilon$ that
starts at $y=0$ and moves upwards with constant velocity as the line sweep
intersects $DOWN_{ij}$. It remains constant at a maximum height as $\ell$
passes thru $MID_{ij}$ and moves downwards to 0 as $\ell$ passes through
$UP_{ij}$. 

This suggests the following operations on the data structures, using
\lemref{zzz2}.  Consider the rectangle $b_j$ of the vertical decompostion
of $L_i$, (which corresponds to translations for which $a_i$ is in the
vicinity of $b_j$).  We divide $b_j$ into three rectangles $b_{ij, UP}$,
$b_{ij,MID}$ and $b_{ij,DOWN}$, which are the intersection regions of
$b_j$ and $ UP_{ij}$, $MID_{ij}$ and $DOWN_{ij}$.  As the linesweep hits
the upper boundary of a rectangle $b_{ij, UP}$, we insert the moving
segment $u_{ij}(\tau)$ into $T(\tau)$. When $\ell$ reaches the upper
boundary of $b_{ij,MID}$ we insert a horizontal moving segment
$u'_{ij}(\tau)$ chosen such that that $u_{ij}(\tau) + u'_{ij}(\tau)$
equals $Max_{ij}$. This is done in order to avoid deleting or changing
$u_{ij}(\tau)$. When $\ell$ reaches the upper boundary of $b_{ij,DOWN}$,
we insert into $T(\tau)$ the segment $u''_{ij}(\tau)$ which is also
decreases linearly as $\tau$ decreases, and is choosen such that
$u(\tau)_{ij}+ u'(\tau)_{ij} + u''_{ij}(\tau)$ equals $\Cov(a_i, t+b_j)$
at this translation $t$, $t\in DOWN_{ij}$. Overall, we add three (moving)
segments for each rectangles of $L_i$, and since the number of these
rectangles is $O(n^2)$, it follows that the overall running time of the
algorithm is $O(n^3\log^2n)$. Note also that at each update, we decrease the
current ``time'' $\tau$; this is a constant time operation per update. 
\end{pf}

\subsection{Maximum coverage for horizontal segments }\subseclab{horizontal} 

This is a line-sweep algorithm reminiscient of 
the  Chew-Kedem~\cite{ChewKedem} and Chew \emph{et al.}~\cite{CDEK}
algorithm for computing the similarity between point-sets in the plane,
under the $\ell_\infty$ norm. As in~\subsecref{horizontal+vertical}, 
we define layers $L_i$ for each endpoint $p_i$ of segments in $A$.
Construct a 
horizontal decomposition of $L_i$, breaking it into a collection
$\B_i=\{\beta_{i1}\, \beta_{i2}\dots \}$ of $O(n)$ interior-disjoint
rectangles.

Let $\S$ denote the set of {\em vertical} segments on the boundaries of
the layers $L_i$ (for $i=1\dots 2n$).  Let $\T$ be a segment tree
constructed on the 
segments of $\S$.  During the algorithm we sweep the translation plane
\TP\ using a vertical sweep line $\ell$.  Once $\ell$ meets a segment
$e\in \S$, we insert $e$ into $\T$.  No segment is deleted. 

Let $\mu$ be a node of $\T$. Let $I_\mu$ be the horizontal infinite strip
whose $y$-span is the interval of $\mu$ and let $S_\mu\subseteq\S$ denote
the segments on or to the left of $\ell$ which correspond to
$\mu$  i.e. the segments whose $y$-span contains $I_\mu$ but  not
 $I_{father(\mu)}$.  We maintain the following fields with each
node $\mu$ of $\T$. All of these are set to zero at the beginning of the
algorithm:

\begin{itemize}    
\item  $last_\mu$: the last $x$ event at which a segment was inserted into $S_\mu$. 
\item $\Pos_\mu $: the number of segments in $S_\mu$ resulting from the
 right (resp.\ left) endpoint of a segment $a\in A$ meeting a left (resp.\
 right) vertical segment of some layer.  We call such an event a {\em
 Positive event} 
\item $\Neg_\mu$: the number of segments in $S_\mu$ resulting from the
left (resp.\ right) endpoint of a segment $a\in A$ meeting a left (resp.\
right) vertical segment of some layer.  We call such an event a {\em
Negative event}. 
\item $w_\mu$: The maximal coverage obtained by segments stored at $S_\mu$
itself. 
\item  $\Cov_\mu$: The maximal coverage obtained by events of ``segments'' 
 stored at the descendants nodes of $\mu$ {\em including} $\mu$ itself.
\end{itemize} 

\noindent{\bf Performing an insertion: } 
Once  $\ell$ hits a new segment $s\in \S$, we first find all nodes $\mu$
for which $s \in S_\mu$  as in a standard segment
tree. Next, for each such node $\mu$, we increase either $\Pos_\mu$ or
$\Neg_\mu$ by one, according 
to the type of $s$. Next  we add to $w_\mu$ the quantity $( \Pos_\mu
-\Neg_\mu)d$, where $d$ is the horizontal distance from the previous
insertion event into $S_\mu$, (stored at $last_\mu$) till the current
position of the $\ell$.  
We  update $\Cov_\mu$  for each $\mu$ in bottom-up fashion, namely: $ \Cov_\mu = \max \{ \Cov_{left( \mu )}, \Cov_{right(\mu)} \} + w_\mu $.
Each insertion can be performed in $O(\log n)$ time, 
so the overall running time of the algorithm is $O(n^2\log n)$.
When the algorithm terminates, we report a translation $t_{output}$ that
corresponds to the maximum value of $\Cov_{root(\T)}$ obtained by the
algorithm.

\noindent{\bf Remark: } The algorithm can easily be modified to handle the
{\em weighted} case, where  each segment has a weight, and the
contribution to the coverage of a segment is the length  of the covered
portions times the weight of the segment. This is useful when some
segments are more important than others.  

\begin{theorem} 
 Let $t^*\in \TP$ be the leftmost translation that maximises $\Cov(t+A,
 B)$.  Then when the line-sweep passes through $t^*$, $(t^*+A, B) =
 \Cov_{root(\T)}$.
  \end{theorem} 
\begin{pf}
We first make the following observation.  Consider the infinite horizontal
ray $r$ emerging from $t^*$ to the left.  Let $x_1\dots x_l$ be the
$x$-coordinates of the events encountered along this ray, ordered from
left to right.  Let $\Pos_i$ (resp.\ $\Neg_i$) be defined as the number of
positive intersection points of $r$ to the left of $x_i$, with boundaries
of layers that corresponds to positive (resp. negative) events, as
described above.  Clearly

\begin{equation}\eqlab{1} 
\Cov(t^*A, B) = \Sigma_{ i=1 } ^l (\Pos_{i} - \Neg_{i}) (x_{i} - x_{i-1} )
\end{equation}

On the other hand, the sum of the right hand side of \eqref{1} equals
 the sum of the fields $w_\mu$, taken over all nodes $\mu$ of the
segment tree on the path from the root to the leaf node containing $t^*$,
at the instance when the line sweep intersects $t^*$.  This follows from
the fact that each event $x_i$ is also an event in one of the nodes $\mu$
along this path.  
Therefore this sum  equals $\Cov_{root(\T)}$, since the sum of  
the fields $w_\mu$ along every path from the root to a leaf equals
$\Cov(t+A, B)$ at any translation $t$ stored at that leaf, and $t^*$ by
our assumption is maximal.  
\end{pf}

%----------------
%\begin{pf} 
%We first make the following observation.  Consider the infinite horizontal
%ray emerging from $t^*$ to the left.  Let $x_1\dots x_l$ be the
%$x$-coordinations of the events encountered along this ray. Let $\Pos_i,
%\Neg_i$ be defined as the number of positive and negative events at $x_i$
%(for $i=1\dots l$)
 
%Let $a\in A$, and let $x_{i_1}\dots x_{i_k}$ be the events in which an
%endpoint of $a$ is involved.  Clearly for each segment $a\in A$ the total
%portion of $a$ which is covered (under the translation $t^*$) is equal to
% \begin{equation}\eqlab{2} 
%    \Sigma_{ j=1 } ^k (\Pos_{i_j} - \Neg_{i_{j-1}}) (x_{i_j} -
%    x_{i_{j-1}})
% \end{equation} 
%Let $\pi$ be the path of $\T$ from the root toward the leaf containing
%$t^*$ at the instant where the sweep line contains $t^*$.  Summing the fields
% $w_\mu$ over all nodes $\mu$ along $\pi$, and for all events determined by
% segments of $A$, we obtain
% \begin{equation} \eqlab{1} 
%    \Cov(t^*+A, B) = \Sigma_{i=2}^l (\Pos_i-\Neg_i) (x_i-x_{i-1})
% \end{equation} 
% This sum can be seen to equal $\Cov_{root(\T)}$.
%\end{pf} 
%----------------------
\subsection{A lower bound}

Rucklidge \cite{Rucklidge-lower-bound} showed that given a parameter
$\eps$ and two families $A$ and $B$ of segments in the plane, the
combinatorial complexity of the regions in the translations plane (\TP) of
all translations $t$ for which $h(t+A, B)\leq \eps$ is in the worst case
$\Omega(n^4)$, where $h(A,B)$ is the one way Hausdorff distance from $A$
to $B$. 
We show that the $\Omega(n^4)$ bound holds even in the case that all  segments are
\emph{horizontal} (the proof is deferred to Appendix~\ref{lowerbound}). This implies:

\begin{theorem}
\theolab{lowerbound}
The region of all translations $t$
for which $Cov(A, t+B)$ is maximal has combinatorial complexity  $\Omega(n^4)$.
\end{theorem}

\mysec{ Matching Horizontal Segments Under Vertical
Translation}\seclab{vertical-translation} 
In this section we describe a
sub-quadratic algorithm for the Hausdorff matching between sets $A$ and $B$
of horizontal segment, when translations are restricted to the vertical
direction. 

Let $\rho^* = \min_t h(t+A, b)$ where $t$ varies over all vertical
translations, and $h(\cdot, \cdot)$ is the one-way Hausdorff distance. Let
$M$ denote the ratio of the diameter to the closest pair of segments in
$A\cup B$. Further, let $[M]$ denote the set of integers $\{1\dots M\}$.

\begin{theorem}\theolab{horizontal+vertical} 
Let $A$ and $B$ be two set of
horizontal segments, and let $\eps<1$ be a given parameter. Then we can
find a vertical translation $t$ for which $h(t+A, B)\leq (1+\eps)
\rho^*$ in  time  $O(n^{3/2} \mbox{poly}( \log M, \log n,
1/\epsilon))$.
\end{theorem} 

We first relate our problem to a problem in string matching:

\begin{Def}{\bf (Interval matching):} given two sequences $t=t[1] \ldots t[n]$ 
   and $p=p[1] \ldots p[m]$, such that $p[i] \in [M]$ and $t[i]$ is a
   union of disjoint intervals $\{a_i^1 \ldots b_i^1\} \cup \{a_i^2 \ldots
   b_i^2\} \ldots$ with endpoints in $[M]$, find all translations $j$ such
   that $p[j] \in t[i+j]$ for all $i$.  The {\em size} of the input to
   this problem is defined as $s=\sum_i |t[i]|+m$.
\end{Def} 

We also define the {\em sparse} interval matching problem, in which both
$p[i]$ and $t[i]$ are allowed to be equal to a special empty set symbol
$\emptyset$, which matches any other symbol or set.  The size $s$ in this
case is defined as $\sum_i |t[i]|$ plus the number of non-empty pattern
symbols. Using standard discretization techniques~\cite{CS,IMV}, we can show that 
the problem of $(1+\epsilon)$-approximating the minimum Hausdorff distance
between two sets of $n$ horizontal intervals with coordinates from $[M]$
under vertical motion can be reduced to solving an instance of sparse
interval matching with size $s = O(n)$. 
 
Having thus reduced the problem of matching segments to an instance of
sparse interval matching, we show that:

  \noindent $\bullet$ The (non-sparse) interval matching problem can be solved in time
 $O(s^{3/2}\polylog s)$. 

\noindent $\bullet$ The same holds even if the pattern is allowed to consists of unions of intervals. 

\noindent $\bullet$ The sparse interval matching problem of size $s$ can be reduced to
  $O(\log M)$ non-sparse interval matching problems, each of size
  $s'=O(s\ \polylog s)$.  

%\begin{itemize} 
%\item The (non-sparse) interval matching problem can be solved in time
% $O(s^{3/2}\polylog s)$. 
%\item The same holds even if the pattern is allowed to consists of unions of intervals. 
%\item The sparse interval matching problem of size $s$ can be reduced to
%  $O(\log M)$ non-sparse interval matching problems, each of size
%  $s'=O(s\polylog s)$.  
%\end{itemize} 

These three observations yield the proof
of~\theoref{horizontal+vertical}. In the remainder of this section, we
sketch proofs of the above observations.

\noindent{\bf The interval matching problem.} Our method follows the
approach of~\cite{Ab,Ko} and~\cite{AF}; therefore, we  sketch the
algorithm here, omitting detailed proofs of correctness.

Firstly, we observe that the universe size $M$ can be reduced to $O(s)$,
by sorting the coordinates of the points/interval endpoints and replacing
them by their rank, which clearly does not change the solution.  Then we
reduce the universe further to $M'=O(\sqrt{s})$ by merging some
coordinates, i.e. replacing several coordinates $x_1 \ldots x_k$ by one
symbol $\{x_1 \ldots x_k\}$, in the following way.  Each coordinate (say
$x$) which occurs more than $\sqrt{s}$ times in $t$ or $p$ is replaced by
a singleton set $\{x\}$ (clearly, there are at most $O(\sqrt{s})$ such
coordinates).  By removing those coordinates, the interval $[M]$ is split
into at most $O(\sqrt{s})$ intervals.  We partition each interval into
smaller intervals, such that the sum of all occurrences of all coordinates
in each interval is $O(\sqrt{s})$.  Clearly, the total number of intervals
obtained in this way is $\sqrt{s}$.  Finally, we replace all coordinates
in an interval by one (new) symbol from $[M']$ where $M'=O(\sqrt{s})$.  By
replacing each coordinate $x$ in $p$ and $t$ by the number of a set to
which $x$ belongs, we obtain a ``coarse representation'' of the input,
which we denote by $p'$ and $t'$.

In the next phase, we solve the interval matching problem for $p'$ and
$t'$ in time $\tilde{O}(nM')$  using a Fast Fourier Transform-based
algorithm (see the above references for details).  Thus we exclude
all translations $j$ for which there is 
$i$ such that $p[i]$ is not included in the {\em approximation} of $t[i+j]$.
However, it could be still true that $p[i] \notin t[i+j]$ while $p'[i] \in
t'[i+j]$.  Fortunately, the total number of such pairs $(i,j)$ is bounded
by the number of new symbols (i.e.  $M'$) times the number of pairs of all
occurrences of any two (old) symbols corresponding to a given new symbol
(i.e. $O(\sqrt{s}^2)$).  This gives a total of $O(s^{3/2})$ pairs to
check.  Each check can be done in $O(\log n)$ time, since we can build a
data structure over each set of intervals $t[i]$ which enables fast
membership query.  Therefore, the total time need for this phase of the
algorithm is $\tilde{O}(s^{3/2})$, which is also a bound for the total
running time.

The generalization to the case where $p[i]$ is a union of intervals
follows in  essentially the same way, so we skip the description here.

\noindent 
{\bf The sparse-to-non-sparse reduction.} 
The idea here is to map the input sequences to sequences of length $P$,
where $P$ is a random prime number from the range $\{c_1 s \log M \ldots
c_2 s \log M\}$ for some constants $c_1,c_2$.  The new sequences $p'$ and
$t'$ are defined as $p'[i] = \cup_{i': i' mod P=i} \, p[i']$ and
$t'[i]=\cup_{i': i' mod P=i} \, t[i']$.  It can be shown (using
similar ideas as in~\cite{CS}) that if a translation $j$
{\em does not} result in a match between $p$ and $t$, it will remain a
mismatch between $p'$ and $t'$ with constant probability.  Therefore, all
possible mismatches will be detected with high probability by performing
$O(\log M)$ mappings modulo a  random prime.

\comment{

\mysec{Dynamic wrapping } 

The technique of dynamic warping is bla bla. Assume $\alpha$ and $\beta$
are polygonal paths of $n$ edges each, and that that there are $m$ sample 
point on each of them. We show ... $O(mn$, in contrast to the 
known $O(m^2)$.

The idea is to create a map $\M$ where the $x$-axis corresponds 
to $\al$, the $y$-axis corresponds to $\beta$, and the $i,j$-cell  
$c_{ij}$ of the map corresponds to the edge $e_i, e_j$. 

Assume that the ``dog'' is on $\beta$ and that the ``man'' is on $\al$. 
For each point $q$ on the boundaries of the $c_{ij}$ let 
$w_q$ denote the weight of the shortest path leading to $q$. 
Assuming dynamic programming approach, we seek now 
to compute the values on the other sides of $c_{ij}$. 
since the optimal path is both $x$-monotone and $y$-monotone, we 
assume that $q$ lies on the left boundary of $c_{ij}$. 

We assume a model where the distance between two segments is computable in 
time $O(1)$. 

Let $e_l, e_r, e_u, e_d$ denote the left, right up and down 
edges of $c_{ij}$.

Let $a_1\dots a_k$ denote the points of $\beta$ corresponding to the
finer grid on  
$e_l$, and let $b_1\dots b_k$ be the ones on $e_r$ .
Let $f_t$ denote the  optimal path that connects $b_t$, and let $a_{f_t}$
denote the  
left endpoint of $f_t$. 

}

\mysec{Computing The \frs\ Distance Under Translation}\seclab{frechet}
 %\documentclass{article}
%\input{defns}
%\newcommand{\Fr}{Fr\'{e}chet\ }
%\newcommand{\Frd}{Fr\'{e}chet distance}
%\newcommand{\f}{\ensuremath{d_F}}
%\newcommand{\wf}{\ensuremath{d_{\tilde{F}}}}

%\title{The \Frd}
%\begin{document}

In this section, we present algorithms for computing the \Frd between two
polygonal chains. Recall that the \Frd between two curves $P$ and $Q$, $d_F(P 
,Q)$ is defined as: 
\[ d_F(P,Q) = \inf_{\alpha,\beta} \max_{t \in [0,1]} \|f(\alpha(t))-g(\alpha(t))\| \] 
where $\alpha,\beta$ range over continuous increasing functions from
$[0,1]\rightarrow [a,a']$ and $[0,1]\rightarrow [b,b']$ respectively.

Dropping the restriction that $\alpha, \beta$ are increasing functions
yields a measure we call the \emph{weak} \Fr distance, denoted by
$\wf$. It can be easily seen that both $\f$ and $\wf$ are metrics.

Let the curves $P$ and $Q$ be length-parameterized by $r, s$. In other
words, $P = P(r), Q = Q(s)$, where $0 \le r,s \le 1$. For any fixed
$\epsilon$, let 
$F_\epsilon(P, Q)$, the \emph{free space}, be defined as
\[ F_\epsilon(P,Q) = \{(r,s) \mid \|P(r)-Q(s)\| \le \epsilon \}\]
where $\|\cdot\|$ is the underlying norm\footnote{In this section, we will
  consider the $l_2$ norm unless otherwise specified.}. The free space
captures the space of parameterizations that achieve a \Frd of at most
$\epsilon$. In the sequel we will denote the free space by $F_\epsilon$
when the parameters $P$ and $Q$ are clear from the context.

Let a  polygonal chain $P : [0,n] \rightarrow
\reals^2$ be a curve such that for each $i \in \{0,\ldots,n-1\}$,
$P_{|[i,i+1]}$ is affine i.e 
$P(i+\lambda) = (1-\lambda)P(i)+\lambda P(i+1), 0 \le \lambda \le 1$. 
For such a chain $P$, denote $|P| = n$. Let
$P_i$ denote the segment $P_{|[i,i+1]}$. For two polygonal chains $P, Q$
where $|P|=p, |Q|=q$, and a fixed $\epsilon$, the free space $F_\epsilon \subseteq 
  [0,p]\times [0,q]$ is given (as before) by:
\[ F_\epsilon(P,Q) = \{(r,s) \mid \|P(r)-Q(s)\| \le \epsilon \}\]

Let $F^{ij}_\epsilon = F_\epsilon \cap (P_i\times Q_j)$. 
Observe that $F^{ij}_\epsilon = F_\epsilon(P_i, Q_j)$. It can be
seen~\cite{AG92} that $F^{ij}_\epsilon$ is the affine inverse of a unit
ball with respect to the underlying norm. Consequently, $F^{ij}_\epsilon$
is convex.  

Consider the points of intersection of a single cell $C_{ij} =
F^{ij}_\epsilon$ with the line segment from
$(i,j)$ to $(i,j+1)$. Since $C_{ij}$ is convex, there are at most two such
points, 
which we denote as $a_{ij}, b_{ij}$, where $a_{ij}$ is below
$b_{ij}$. Similarly, let $c_{ij}$ and $ d_{ij}$ be the points of
intersection of $C_{ij}$ with the line segment from $(i,j)$ to $(i+1,j)$,
where $c_{ij}$ is to the left of $d_{ij}$.

\begin{figure}[htbp]
  \begin{center}
    \psfig{figure=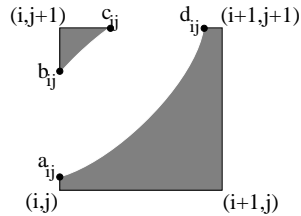,width=1.5in}
    \caption{A single cell in the free space}
    \label{fig:cell}
  \end{center}
\end{figure}
We define an order on the points as follows: For any two points 
$p_1 = (x_1, y_1), p_2 = (x_2, y_2)$, $p_1 \le p_2$ if $x_1 \le x_2$ and 
$y_1 \le
y_2$.

Let an $(x,y)$-monotone path be a path that is increasing in both $x$ and 
$y$ coordinates. Alt and Godau~\cite{AG92} observed that the existence of
a $(x,y)$-monotone path in $F_\epsilon$ from $(0,0)$ to $(p,q)$ is a
necessary and sufficient condition for $\f(P, Q) \le \epsilon$. A similar
property holds for $\wf$; namely, the existence of \emph{any}
non-self-intersecting path in $F_\epsilon$ from $(0,0)$ to $(p,q)$ implies
that $\wf(P, Q) \le \epsilon$. Denote the property ``$(p,q)$ is reachable
from $(0,0)$'' as property ${\cal P}$ (similarly define $\tilde{\cal P}$).

We wish to solve a decision problem for  the \Frd between $P$ and $Q$ minimised over
translations i.e given $\epsilon$, we wish to check whether
$ \min_{t} \f(P, Q+t) \le \epsilon $

\paragraph{The configuration space}
A \emph{critical event} is one that can change the truth value of ${\cal
  P}$. Each such event is one of the following two types:
(1) The intersection points $a_{ij}, b_{ij}, c_{ij}, d_{ij}$ appear (or
  disappear).
(2)  For two cells $C_{ij}$ and $C_{kj}, k > i$, $a_{ij}$ and $a_{kj}$
  (or $b_{kj}$) 
  change their relative vertical ordering. Analogously, for two
  cells $C_{ij}$ and $C_{ik}, k > j$ the  points $c_{ij}$ and
  $c_{ik}$ (or $d_{ik}$) change their relative horizontal ordering.

Type 2 events correspond to the creation or deletion of \emph{tunnels}. 
For any point $r$ in the space $[0,p]\times [j,j+1]$,
let $k$ be the \emph{rightmost} interval such that 
$r$ projected onto the interval $[a_{kj},b_{kj}]$ lies
between the endpoints of the interval. We define \emph{rt}
$(r) = k$. For any point 
$r \in [i,i+1]\times[0,q]$, let $k$ be the \emph{topmost} interval such that 
$r$ projected onto the interval $[c_{ik},d_{ik}]$ lies  
between the endpoints of the interval. We define\footnote{The term
  \emph{rt} denotes a \emph{right tunnel}; \emph{ut} denotes an
  \emph{upper tunnel}.} \emph{ut}$(r) = k$.  

As $Q$ translates, each of the $x_{ij}, x \in \{a,b,c,d\}$ can be
represented as a function $x_{ij}(t) : \reals^2 \rightarrow [0,1]$.
\begin{prop}
\label{prop:cspace}
For a point $x_{ij}$, the function $x_{ij}(t)$ is a second degree
polynomial in the coordinates of $t$. 
\end{prop}

\paragraph{From free space to a graph}

Our algorithm for computing $\f(P,Q)$ is based on a reduction of the
problem to a directed graph reachability problem. Intuitively, we can
think of a monotone path in the free space as a  path in a directed graph
(actually a DAG). The advantage of this approach is that
we can exploit known methods for maintaining graph properties dynamically
in an efficient manner. Thus, as we traverse the space of translations, we
need not recompute the free space at each critical event.

Let  
$V = \bigcup_{i,j}\{ v^a_{ij}, v^b_{ij}, v^c_{ij}, v^d_{ij} \}$ and 
$ T = \bigcup_{i,j,i <  k \le p}\{ t^a_{ijk}, t^b_{ijk} \} \ \ \cup
\ \bigcup_{i, j, j <  k \le q}\{ t^c_{ijk}, t^d_{ijk} \} $
where $0 \le i \le p$ and $0 \le j \le q$.
The vertices in $V \cup T$ are associated with points of the free
space. More precisely, vertex $v^x_{ij}$  is
associated with the point $x_{ij}$ (where $x$ is one of $\{a, b, c,
d\}$). Vertex $t^x_{ijk}$ is associated with the projection of point $x_{ij}$
onto the interval $[a_{kj},b_{kj}]$ ($x \in \{a, b\}$), and vertex
$t^y_{ijk}$ is associated with the projection of point $y_{ij}$ 
onto the interval $[c_{ik},d_{ik}]$ ($y \in \{c, d\}$). We define $f(v) =
p$, where $p$ is the point associated with vertex v. 

Let $V^1_{ij} = \{ v^a_{ij}, v^b_{ij} \} \cup \bigcup_{l < i \le
  rt(a_{lj})} t^a_{lji} \cup \bigcup_{l < i \le rt(b_{lj})} t^b_{lji}$ and
 $V^2_{ij} = \{ v^c_{ij}, v^d_{ij} \} \cup \bigcup_{l < j \le ut(c_{il})}
t^c_{ilj} \cup \bigcup_{l < j \le ut(d_{lj})} t^d_{ilj}$. 
$V^1_{ij}$ denotes the set of vertices associated with points on the line
segment from $(i,j)$ to $(i,j+1)$.  Similarly, $V^2_{ij}$ denotes the set
of vertices associated with points on the line segment from $(i,j)$ to
$(i+1,j)$. In addition, $V^1_{ij}$ and $V^2_{ij}$ contain vertices
associated with points whose \emph{tunnels} cross the cell $C_{ij}$. 

We now describe the construction of the edge set for each $(i,j)$. 
Firstly, set $E^1_{ij} = \{(v, v^b_{ij}) \mid v\in V^1_{ij} \}$ and set 
$E^2_{ij} =\{(v, v^d_{ij}) \mid v\in V^2_{ij} \}$ 
For each $v \in V^1_{ij}$, let $n(v) = \arg \min_{v' \in V^1_{i+1,j},
  f(v') \ge v} f(v')$
 Similarly, for each $v \in V^2_{ij}$, let $n(v)$ denote the vertex
in $V^2_{i,j+1}$  having the same property. Let 
$E^3_{ij} = \{ (v,n(v)) \mid v \in V^1_{ij} \cup V^2_{ij} \}$. 
%Notice that as a consequence of the above definition, $n(t^a_{ijk}) =
 %t^a_{ij,k+1}$ for all $i,j,k$. A similar property holds for $t^x_{ijk},
 %x = \{ b,c,d\}$.  
Finally, set $E^4_{ij} = \{ (v^b_{ij}, v^c_{i,j+1}),(v^d_{ij},
v^a_{i+1,j})\}$. Now, we set $E_{ij} = E^1_{ij} \cup E^2_{ij} \cup
E^3_{ij} \cup E^4_{ij}$.  

Let $E = \bigcup_{i,j} E_{ij}$. 
This yields the directed graph $G = (V\cup T, E)$. Note that $|V\cup T| =
O(pq(p+q))$ and $|E| = O(pq(p+q))$. Also, it is easy to see that for any
edge $(u,v) \in E$, the straight line from $f(u)$ to $f(v)$ is an
$(x,y)$-monotone path. 
We first show that reachability in the graph $G$ is equivalent to 
path construction in $F_\epsilon$. The proof of this theorem is
straightforward and is deferred to Appendix~\ref{app:graph}.

\begin{theorem}
\label{thm:graph-free}
An $(x,y)$-monotone path from $(0,0)$ to $(p,q)$ exists in $F_\epsilon$
iff $v^b_{pq}$ is reachable from $v^a_{00}$ and $f(v^a_{00})=(0,0),
f(v^b_{pq})=(p,q)$.
\end{theorem}

For every edge $e \in E$, let $\gamma(e) \subseteq \reals^2$ be the set of
translations $t$ such that in the graph $G$ constructed from the free
space $F_\epsilon(P,Q+t)$, the edge $e$ is present. Let $\Gamma$ be the
arrangement of all the $\gamma(e)$. We first establish a bound on the
complexity of $\Gamma$.

The following three propositions (which we state without proof), follow
from Proposition~\ref{prop:cspace}. Roughly speaking, with each edge $e$
we can  associate a boolean combination of predicates $P_1, P_2, \ldots, P_k$, where each
predicate compares some constant degree polynomial to zero. (i.e the
regions are semi-algebraic sets). 

\noindent $\bullet$ For any region $\gamma(e)$, the boundaries consist of
segments of curves described by constant degree polynomials. 

\noindent $\bullet$ For an edge $e \in E_{ij} -T\times T$, the region
$\gamma(e)$ is a constant number of simple regions of constant description
complexity. 

\noindent $\bullet$ For an edge of the form $(t^x_{ijk},t^x_{ijk+1}), x
\in \{a,b,c,d\}$, the region $\gamma(e)$ consists of a set of simple regions of total
description complexity $k$. 

\begin{lemma}
$|\Gamma| = O(p^2q^2(p+q)^4)$.
\label{lemma:complexity}
\end{lemma}
\begin{proofsketch}
There are $O(pq(p+q))$ edges. For each edge $e$, the complexity of the
associated region can be at most $O(p+q)$. Since any pair of constant
degree polynomials intersect in a constant number of points, the overall
complexity of $\Gamma$ is given by $(pq(p+q)\times(p+q))^2$.
\end{proofsketch}

\begin{lemma}
Let $\gamma_k = \gamma((t^x_{ijk},t^x_{ij,k+1}))$, where $x \in
\{a,b,c,d\}$. Then for all $l$ such that $i \le l < k$, 
$\gamma_k \subseteq \gamma_l$.
\label{lemma:tunnel-nesting}
\end{lemma}
\begin{pf}
Whenever the edge $(t^x_{ijk},t^x_{ijk+1})$ is present, all
edges of the form $(t^x_{ij,l},t^x_{ijl+1}), i \le l < k$ must also be
present. 
\end{pf}

Theorem~\ref{thm:graph-free} indicates that the graph property that we
need to maintain is the reachability of $v^b_{pq}$ from $v^a_{00}$. 
The algorithm is now as follows: Fix a traversal of the arrangement of
regions. Check reachability at the starting cell. Each time an
edge is crossed in the traversal, it corresponds to the deletion (and
insertion) of edges in the graph, which we use to update the graph and
check for reachability. Stop whenever the above property holds, returning
YES, else return NO. 

\begin{theorem}
\label{thm:graph-space-dynamic}
Iff there exists a translation $t$ such that $\f(P,Q+t) \le \epsilon$, the
above algorithm will terminate with a YES. 
\end{theorem}
\begin{pf}
Consider a type 1 critical event, where the interval $a_{ij},b_{ij}$ is
created. This interval corresponds to the edge
$(v^a_{ij},v^b_{ij})$. Hence, this event corresponds to entering the
region associated with the above edge. Similar arguments hold for other
type 1 critical events. 

Suppose we have a type 2 critical event, where the point $a_{kj}$
rises above $a_{ij}$ (in their relative vertical ordering). Note that this event does
not change the reachability of $(p,q)$ in the free space unless
rt$(a_{ij}) > k$. If this is the case, then the event results in
setting rt$(a_{ij}) = k$, implying that all edges of the form
$(t^a_{ijl},t^a_{ij,l+1}), l \ge k$ are deleted, which corresponds to
leaving the regions corresponding to this set of edges\footnote{Note that since
the regions corresponding to this set of edges are nested (by
Lemma~\ref{lemma:tunnel-nesting}), such a transition is indeed
possible. In fact, the existence of such a critical point implies that all
of these regions intersect in at least one point that is also contained in
$r((t^a_{ij,k-1},t^a_{ijk}))$. The critical event can be interpreted as
the result of the translation across this point.}. 

Conversely, it can be seen that any transition from one cell of the
arrangement to another corresponds to a critical event. We defer the
details to a full version of the paper. 
\end{pf}

It now remains to analyse the complexity of the above algorithm. A
transition between cells yields $O(1)$ updates, except in the case
described in Theorem~\ref{thm:graph-space-dynamic} above, where a transition
occurs across the boundary of region $r((t^a_{ij,l-1},t^a_{ijl}))$ into
the region $r((t^a_{ij,k-1},t^a_{ijk}))$, causing $\Theta(l-k)$
updates. However, note that in this event, it must be the case that all
the regions $r((t^a_{ij,m},t^a_{ij,m+1}), k \le m < l-1$ intersect at this
transition point (from Lemma~\ref{lemma:tunnel-nesting}), and thus the
cost of this transition can be distributed 
among these cells. Hence, the total number of updates is given by
Lemma~\ref{lemma:complexity}. 

%Na\"{\i}vely implemented, each update takes time $O(pq(p+q))$ (use a depth
%first search to check reachability), yielding an overall running time of
%$O(p^3q^3(p+q)^4)$.  
%However, an improved result can be obtained by using the method of
%cuttings~\cite{C93}. 
%The idea is to construct a cutting over the arrangement,
%so that we decompose it into \emph{super-cells} where each such super-cell
%intersects only a few of the regions. Thus in each super-cell, there are a
%small number of edges that participate in updates. We will show that given
%a directed graph $G$ with $n$ vertices and $m$ edges, but only  $k$ edges that
%participate in updates, reachability can be maintained over updates in
%time $O(\min(mk,n^\omega) + k^2U)$, where $U$ is the number of updates, thus saving over
%the na\"{\i}ve algorithm that runs in time $O(\min(mk,n^\omega)U)$. This
%technique is similar to the method employed in~\cite{KMY98}
% in the context of lookahead for dynamic graph algorithms. 

To determine reachability, we must now traverse the arrangement. For ease
of notation, we will assume that $p = \Theta(q)$ and set $n = p+q$. 
The arrangement
consists of $O(n^3)$ regions, each described by $O(n)$ curves of constant
description complexity. Let us fix $r$ (we will specify the value of $r$
later). It can be shown (using the theory of
cuttings~\cite{C93,BS95}) that we can compute a subset ${\cal R}$ of the
regions of size $O(r\log r)$ with the property that if we compute the
vertical decomposition of each \emph{super-cell} in the arrangement of
${\cal R}$, each of the resulting \emph{primitive super-cells} (of constant
complexity) is intersected by $O(n^3/r)$ regions. 

\begin{lemma}
Given a graph $G = (V,E), |V|= N, |E|=M$, designated nodes $s,t \in V$,
and a set of $k$ edges $E' \subset E$, $s$-$t$ reachability in $G$ can be
maintained over edge insertions and deletions from $E'$ in total time
$O(\min(N^\omega, Mk)+ k^2U)$, where $U$ is the number of such updates
($\omega$ is the exponent for matrix multiplication). 
\label{lemma:supercell}
\end{lemma}
\begin{pf}
Let $V'$ be the set of endpoints of edges in $E'$. We
compute the graph $G' = (V'' = V' \cup \{s,t\}, E'')$, where $(u,v) \in
E''$ if there is a directed path from $u$ to $v$ in $G$. Note that $|V''|
\le 2k$. The computation of this graph can be done by performing a full
transitive closure on $G$ that takes time $O(n^\omega)$. Alternatively, we
can perform $O(k)$ depth-first searches (one from each vertex in $V''$) to
construct $G'$. 

Now, to process updates, we update the graph using a standard dynamic
update procedure that takes time $O(k^2\log k)$ time (amortized) per
update\cite{K99}, yielding the result. 
\end{pf}

The algorithm now proceeds as follows: Each primitive super-cell
has a set of edges associated with it
(one for each region that intersects it). We use the above lemma to
perform an efficient dynamic reachability test for each cell of the
original arrangement in this primitive super-cell. When we move to the next
primitive super-cell, we recompute the induced graph and repeat the process.

We now compute the value of $r$. The total number of cells in the
arrangement is $O(n^8)$ by Lemma~\ref{lemma:complexity}. There are
$O(r^2n^2\log^2r)$ primitive super-cells, each intersected by $O(n^3/r)$
regions. Consider a single  primitive super-cell $i$. We apply
Lemma~\ref{lemma:supercell} with $N = M = O(n^3)$, $k = O(n^3/r)$, and $U
= U_i$, where $U_i$ is the number of cells in $i$. The current value of
$\omega$ is approximately $2.376$~\cite{CW90}, and thus $\min(N^\omega,Mk)
= Mk = n^6/r$ for all $r = \Omega(1)$. The cost of processing $i$ is
therefore $n^6/r$ + $n^6U_i/r^2$. Summing over all primitive super-cells,
and replacing $\Sigma U_i$ by $O(n^8)$, we obtain the overall running time
of the algorithm to be $O(n^8r\log^2r + n^{14}/r^2)$. Balancing, we obtain
an overall running time of $O(n^{10}\polylog n)$.

\begin{theorem}
Given two polygonal chains $P, Q, |P|=p, |Q|= q$, and $\epsilon > 0$, we
can check if $\f(P, Q) \le \epsilon$ in time $O(n^{10}\mbox{\sl polylog\,}n)$.
\end{theorem}

\paragraph{The weak \Frd}
As described earlier, the weak \Frd (denoted by $\wf$) relaxes the
constraint that the parametrizations employed must be monotone. Note that
for any two curves $P,Q$, the following 
inequality is true:
$d_H(P, Q) \le \wf(P,Q) \le \f(P,Q) $
Also,  by the result of Godau~\cite{G98}, all three measures collapse
to one if both curves are convex. The above inequality is significant
because it suggests that the weak \Frd may serve as a relaxed curve
matching measure with possibly more tractable algorithms. 

As it turns out, this is indeed the case. Our techniques from the
previous algorithm apply here as well, with two key differences. Firstly,
since the paths need not be monotone, we no longer need the concept of a
tunnel, thus reducing the number of critical events that need to be
examined to $O(pq)$. Secondly, the underlying graph is now undirected, and
there are  efficient procedures for maintaining connectivity in an
undirected graph~\cite{HLT98}. We defer details to a full version of the
paper, and summarize the result as:

\begin{theorem}
Given two polygonal chains $P, Q, |P|=p, |Q|= q$, and $\epsilon > 0$, we
can check if $\min_t \f(P, Q+t) \le \epsilon$ in time $O(n^4\mbox{\sl
  polylog\,}n)$, where $n = O(p+q)$. 
\end{theorem}

\paragraph{An approximation scheme}

An $(\eps,\beta)$-approximation (defined by Heffernan and
Schirra~\cite{HS94}) for $\f(P,Q)$ under translations can be obtained from
the following observation: 

\begin{lemma}
Given polygonal chains $P, Q$, let $t$ be the translation that maps the
first point of $Q$ to the first point of $P$. Then $\f(P,Q+t) \le 2d^*$,
where $d^* = \min_{\mbox{translations\ } t} \f(P,Q+t)$. 
\end{lemma}
\begin{pf}
Let $t^*$ be the translation such that $\f(P,Q+t^*) = d^*$. Clearly, the
first point in $Q$ is at most $d^*$ away from the first point of
$P$. Applying the translation $t' = t - t^*$ to $Q$, no point in $Q$ is moved
more than $d^*$ units away from its associated point in $P$. Hence,
$\f(P,Q+t^*+t') = \f(P, Q+t) \le 2d^*$.
\end{pf}
Applying the standard discretization trick in a ball of radius $d^*$
around the first point of $P$, we obtain an
$(\eps,\beta)$-approximation for any $\beta > 0$. Note that this
scheme is very efficient, running  in time $O(n^2\poly(\log n$,
$1/\beta))$. 

%\end{document}

%%% Local Variables: 
%%% mode: latex
%%% TeX-master: "segh"
%%% End: 
 
%\subsection*{Acknowledgements} We would like to thank Helmut Alt, Julien
%Basch, Mikkel Thorup, Carola Wenk and Li Zhang for fruitful discussion. We
%also thank H\'ector H.\ Gonz\'alez-Ba\~nos and Eric Mao for supplying some
%of the pictures in this paper

\def\trd{3{\footnotesize\it rd} } 
\newpage
%\bibliographystyle{plain}
%\bibliography{segh}

%=========%===========================================================
\newpage
\appendix

%**************************************************************
%**************************************************************

%====================================================== 
\section{Proof of \lemref{zzz2}} 

\label{app:pf1} 

\def\maxs{\max(\sum_\S(\cdot))} \def\sum{\mbox{\sl sum}} \def\S{{\cal S}}

%\begin{pf} (of \lemref{zzz} and \lemref{zzz2}). 
   \begin{Def} 
      For a geometric object $R$ let $X(R)$, the $x$-span of $R$, denote
      the interval of the $x$-axis between the leftmost and the rightmost
      point of $R'$, where $R'$ is the orthogonal projection of $R$ on the
      $x$-axis.
   \end{Def}

   \begin{claim}\label{claim:l3} 
      Let $P=\{(x_1,y_1),\dots (x_m, y_m) \}$ be a point set. We can
      construct in time $O(m\log^2 m)$ a data structure for $P$ such that
      given a query segment $s$, the point $(x_k, y_k)$ that maximizes the
      $y$-value of the set $ \{ s(x_i) + y_i ~|~ x_i\in X(s), 1\leq i\leq
      m \} $ can be found in time $O(\log^2 m )$.
   \end{claim} 
   %--------------------- 
   \begin{pf} 
      If $X(P)\subseteq X(s)$, then $(x_k, y_k)$ is clearly a vertex of
      the convex hull of $P$, and once the convex hull is computed, we can
      find $(x_k, y_k)$ in time $O(\log n)$. To answer the query in the
      case that $X(P)$ is not contained in $X(s)$, we construct a sorted
      balanced binary tree $\Psi= \Psi(P)$ on the set $\{x_1\dots x_m\}$.
      For each node $\mu\in \Psi$ let $P_\mu $ denote the points in the
      subtree of $\mu$, and let $X_\mu$ denote the $x$-span of $P_\mu$. We
      construct $C_\mu$, the convex hull of $P_\mu$, for each node $\mu$
      of $\Psi$. Once a query segment $s$ is given, we find a set $U$ of
      $O(\log |P|)$ nodes of $\Psi$ with the property that for each node
      $\mu\in U$, $X_\mu$ is contained in $X(s)$, and in addition, each
      $(x_i, y_i)\in P$ for which $x_i\in X(s)$ appears in exactly one of
      the sets $P_\mu$, for $\mu \in U$.  We perform the query suggested
      by the previous claim on $C_\mu$ for each $\mu\in U$.
   \end{pf}

   Based on Claim~\ref{claim:l3}, we describe the data structure as
   follows. Let $m=|\S|$.  First observe that the maximum must be obtained
   at an endpoint of a segment of $\S$.  We partition $\S$ into $\S_1$ and
   $\S_2$. The set $\S_2$ contains at least $m-\sqrt{m}$ of the segment of
   $\S$. It is updated after $\sqrt m$ insertions or deletion operations
   into/from $\S$ Once it is updated, we explicitly compute the function
   $\sum_{\S_1}(\cdot)$, and construct the data structure
   $\Psi=\Psi_{\S_1}$ of Claim~\ref{claim:l3} for the vertices of the
   graph of $\sum_{\S_1}(\cdot)$.  As easily observed, the complexity of
   the graph of $\sum_{\S_1}(\cdot)$ is $O(m)$, since a vertex of this
   function occurs only at endpoint of a segment of $\S_1$, thus the time
   needed to constuct $\Psi =\Psi_{\S_1}$.
% Let $P$ 
%   be the set of vertices of $\sum_{\S_1}(\cdot)$. We construct the data structure $\Psi_P$. 
   The set $\S_2 = \S \setminus \S_1 $ has cardinality $\leq \sqrt{m}$.
   Each time a segment is inserted (resp.\ deleted) into/from $\S$, it is
   inserted (resp.\ deleted) into/from $\S_1$. Once the size of $\S_1$
   exceeds $ \sqrt{m}$, we set $\S_1$ to be $\S$, construct $\Psi$, and
   empty $\S_2$.
   
   In order to maintain the maximum $\maxs$, we do the following. Once a
   segment is inserted or deleted into $\S_1$, we explicitly compute (the
   graph of) $\sum_\S(\cdot)$ which is piecewise linear of complexity
   $O(\sqrt m)$.  With each segment $e$ of this graph (not to be confused
   with the segments of $\S$) we perform a query in $\Psi_{\S_1}$. The
   maximum obtained is is $\maxs$.
   
   Next we describe the modifications of the data structure needed in the
   case where (some of) the segments of $\S$ move vertially in a constant
   speed with the time parameter $\tau$.  Let $X'=\{x_1\dots x_m\}$ denote
   the $x$-coordinates of the endpoints of the segments of $\S$.  They are
   not time dependent. Let $y(x,\tau)$ denote the $y$-value of the sum
   function at the coordination $x$ at time $\tau$.  Clearly as long as no
   insertions or deletions are taken place in $\S$, $y(x,\tau)$ moves
   (vertically) at a constant velocity. It is well known fact that the
   convex hull of such a set of points can go through $O(m)$ combinatorial
   changes, which we can compute in time $O(m \log m )$. This suggest the
   following modification to the data structure of $\T$ as follows. As
   before, each node $\mu$ is associated as before with the convex hull
   $C_\mu = C_\mu(t)$, but now these convex hulls might change in time.
   However, as argued, the total number of changes they go through is only
   $O(m \log ^2m)$.  The query process remains the same. 
   
\section{Proof of~\theoref{lowerbound}}
\label{lowerbound}
  
Assume for the construction that $\eps=1/2$.  The first component in the
construction (see \figref{lowerbd}) is the set $B_1'$ consisting of $2n$
points, which are 
$$ \{(i,{1/2}-i/n) \mbox{ ~ and ~} (i, -{ 1/ 2}- i/n
-1/4n^2), ~~\mbox{for} i= 1\dots n \}\ .$$
 Thus the $i^{th}$ pair $(i,{1 / 2}-i/n)^+ \mbox{ ~ and ~} (i, -{ 1/ 2} - i/n
 -\delta)^+$ (i.e., the Minkowski sum of these points and the $\ell_infty$
 ball) form two  close vertically aligned squares, where the gap
 between them is of unit width, and of height $1/4n^2$. The $i^{th}$ pair is
 located at distance $i/n$ below the $x$-axis.  We add the segment
 $B_1''$, which is the long horizontal segment between the points $(-n,
 -1/4)$ and $(0,-1/4)$ and the segment $B_1'''$ between $(n,-1/4)$ and
 $(2n,-1/4)$. Let $B_1=B_1'\cup B_1''\cup B_1'''$.

\begin{figure}[htbp] 
   \begin{center} 
      \input{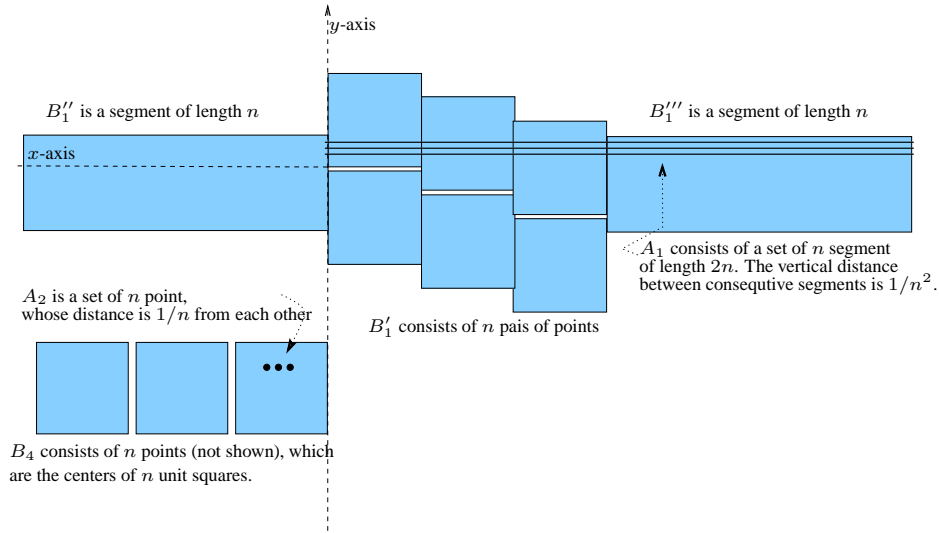}
      \caption{ The lower bound construction for $n=3$. The set $B$ is not
        shown explicitly; only $B^+$ is shown. \figlab{lowerbd} }  
   \end{center} 
\end{figure} 

The set $A_1$ consists of $n$ horizontal segments of length $2n$,
each separated by a gap of $1/n^2$ from the next one. The left endpoint of
all of them is 
on the $y$-axis, and the middle one is on the $x$-axis.  By shifting them
vertically, each segment in turn is not completely covered at some time,
when it passes between the gaps between one of the pairs of $B_1$. In all
other cases, all the segments are completely covered.  The region in \TP\
corresponds to all translations $t$ for which $h(t+A_1, B_1)\leq 1 $
consists of $\Omega(n^2)$ horizontal strips, each of length $n$.

The set $B_2$ consists of the $n$ points $(-(1+1/n^2)i, -5)$ (for
$i=1\dots n$). Thus $B_2^+$ creates $n$ unit squares along the line
$y=-5$, with a gap of $1/n^2$ between them.
The set $A_1$ consist of $n$ points along the horizontal line $(-1/2n,
-5)$ (for $i=1\dots n$).  Observe that $A_1$ fits completely into each of
the squares of $B_2^+$. However, by sliding $A_1$ horizontally, along
$y=-5$ or anywhere at distance $\leq 1$ from $h$, each of the points of
$A_1$ ``falls'' at some stage into each of the gaps between each of the
squares of $B_2^+$, The region $S_2=\{ t \, |\, h(t+A_2, B_2)\leq 1 \}$
consists of $\Omega(n^2)$ vertical strips in \TP, each of hight $2$.
Letting $A=A_1\cup A_2$ and $B=B_1\cup B_2$, the region $S = \{ t \, |\,
h(t+A, B)\leq 1 \}$ is merely the intersection of $S_1$ and $S_2$, which
is clearly of complexity $\Omega(n^4)$, thus proving our claim.
   
\section{Proof of Theorem~\ref{thm:graph-free}}
\label{app:graph}

Suppose $v^b_{pq}$ is reachable from $v^a_{00}$ and 
$f(v^a_{00})=(0,0), f(v^b_{pq})=(p,q)$. Let the path in $G$ be 
$v_1 = v^a_{00}, v_2, \ldots,v_k = v^b_{pq}$.  Replace each vertex $v_i$
by its associated point $f(v_i)$. As observed above, if we now connect the
points $f(v_1), f(v_2), \ldots, f(v_k)$ by straight lines, we obtain an
$(x,y)$-monotone path.

Conversely, suppose there exists an $(x,y)$-monotone path $w$ from $(0,0)$
to $(p,q)$ in $F_\epsilon$. Then $(0,0) \in C_{00}$ and 
$(p,q) \in C_{p-1,q-1}$ and thus $f(v^a_{00}) = (0,0)$ and
$f(v^b_{pq})=(p,q)$. Without loss of generality, we can assume that $w$
consists of a sequence of line segments, where the endpoints of each
segment are one of the $x_{ij}$'s ($x = \{a,b,c,d\}$).

We will show by induction on the number of segments that $v^b_{pq}$ is 
reachable from $v^a_{00}$. Assume that the claim holds for the first $k$
segments on the path. Consider the $(k+1)^{th}$ segment. Let the endpoints
be $w_1, w_2$. By the induction hypothesis, $w_1$ is reachable from
$v^a_{00}$.   

\textbf{Case 1:} Let both $w_1, w_2$ be of the form $x_{ij}, y_{kj}$
respectively, where $x,y \in \{a, b\}$. If rt$(f(w_1)) \ge k$, then the
vertex $t^x_{ijl}$ exists for all $l \le k$, and thus there exists a path
$w_1,t^x_{ij,i+1},\ldots,t^x_{ijk}$. Since $f(t^x_{ijk})$ is on the same
interval as $f(w_2)$ and must be below it, there exists an edge from $t^x_{ijk}$ to $w_2$
in $E_2$ . If on the other hand, rt$(f(w_1)) < k$, 
there must exist one vertex $w' = x^{lj}, i < l < k$ such that $f(w')
> f(w_1)$, and rt$(f(w_1) \le l$. We construct a path from $w_1$ to $w'$
and repeat. 

\textbf{Case 2:} Let both $w_1$ and $w_2$ be of the form $x_{ij}, y_{ik}$
respectively, where $x,y \in \{c, d\}$. An argument similar to Case 1 applies here.

\textbf{Case 3:} Let $w_1 = a_{ij}$ and $w_2 = d_{kl}$. Without loss of
generality we can assume that $k = i$ and $l = j+1$. There exists an edge
from $v^a_{ij}$ to $v^b_{ij})$, which is a predecessor of
$v^c_{i,j+1})$ (using $E_4$), and there exists an edge from
$v^c_{i,j+1})$ to $v^d_{kl}$, thus yielding the desired
path. Other cases can be handled symmetrically. 

Thus, by induction the theorem holds. 

\end{document}